\pdfoutput=1 
 \documentclass[%
 reprint,
 amsmath,amssymb,
 aps,
pre,
]{revtex4-1}

\usepackage{graphicx}
\usepackage{dcolumn}
\usepackage{bm}
\usepackage{subeqnar}
\usepackage{overpic}
\usepackage{color}
\usepackage{adjustbox}
\usepackage[caption=false]{subfig}
\usepackage{rotating}
\usepackage{amsmath}

\begin{document}

\title{Multi-scale Physics of Rotating Detonation Engines: \\Autosolitons and Modulational Instabilities}

\author{James Koch}
\email{jvkoch@uw.edu}
 \affiliation{William E. Boeing Department of Aeronautics and Astronautics, University of Washington, Seattle 98195-2400}
  \author{Mitsuru Kurosaka}
 \affiliation{William E. Boeing Department of Aeronautics and Astronautics, University of Washington, Seattle 98195-2400}
 \author{Carl Knowlen}
 \affiliation{William E. Boeing Department of Aeronautics and Astronautics, University of Washington, Seattle 98195-2400}
\author{J. Nathan Kutz}
 \affiliation{Department of Applied Mathematics, University of Washington, Seattle, WA 98195-3925}

\begin{abstract}
We develop a theoretical framework that predicts and fully characterizes the diverse experimental observations of the nonlinear, combustion wave propagation in a rotating detonation engine (RDE), including the nucleation and formation of combustion pulses, the soliton-like interactions between these combustion fronts, and the fundamental, underlying Hopf bifurcation to time-periodic modulation of the waves. In this framework, the mode-locked structures are classified as autosolitons, or stably-propagating nonlinear waves where the local physics of nonlinearity, dispersion, gain, and dissipation exactly balance. We find that the global dominant balance physics in the RDE combustion chamber are dissipative and multi-scale in nature, with local fast scale (nano- to microseconds) combustion balances generating the fundamental mode-locked autosoliton state, while slow scale (milliseconds) gain-loss balances determine the instabilities and structure of the total number of autosolitons. In this manner, the global multi-scale balance physics give rise to the stable structures - not exclusively the frontal dynamics prescribed by classical detonation theory. Experimental observations and numerical models of the RDE combustion chamber are in strong qualitative agreement with no parameter tuning.  Moreover, numerical continuation (computational bifurcation tracking) of the RDE analog system establishes that a Hopf bifurcation of the steadily propagating pulse train leads to the fundamental instability of the RDE, or time-periodic modulation of the waves. Along branches of Hopf orbits in parameter space exist a continuum of wave-pair interactions that exhibit solitonic interactions of varying strength.
\end{abstract}
\maketitle

\section{INTRODUCTION} \label{intro}
Combustion instabilities are a universal phenomenon in aerospace propulsion systems. In rockets, combustion chambers can exhibit coupling between combustor geometry, propellant injection, and local heat release \cite{CROCCO1962,CLAYTON1968,BENT1968,Zinn1971,Anderson1995} which can lead to instabilities that are capable of inducing mechanical failure \cite{Oefelein1993}, constituting a major risk to the propulsion system.  Historically, to abate this risk, significant resources have been devoted to engineer systems that damp or limit the mechanisms responsible for the formation of instabilities. These engineering tasks are not trivial: the physical processes responsible for the instabilities are highly nonlinear and intricately coupled, often making the unit processes inseparable. Consequently, the physics exploration of these nonlinearities is often constrained to hardware-specific studies. For the Rotating Detonation Engine (RDE), Koch et al. \cite{Koch2020} recently proposed a mathematical model capable of reproducing the diverse, experimentally observed mode-locking dynamics of the RDE.  Here, we build on this model and characterize the fundamental dominant, multiscale balances which drive the instabilities and bifurcation structure in the RDE, showing that the mode-locked states, or autosolitons, are solitonic in how they interact and that the Hopf bifurcation is the fundamental, canonical instability driving bifurcations in the combustion chamber.

\begin{figure*}[]
        \centering
        \begin{overpic}[width=1.0\linewidth]{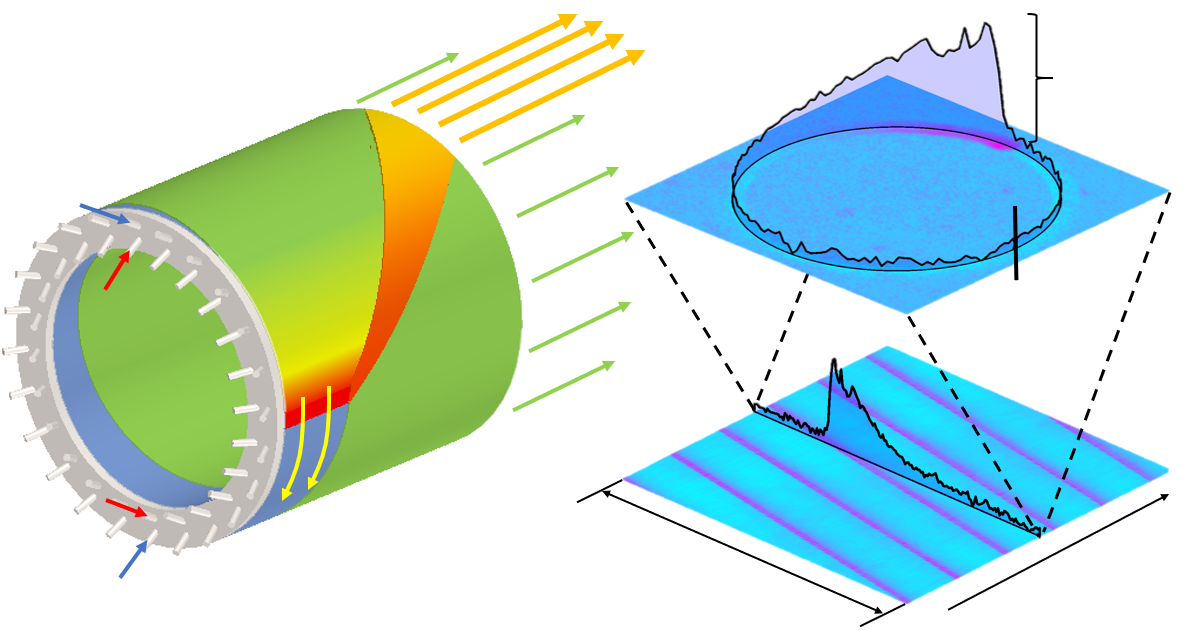}   
        \thicklines
                 
     	\put(57,4){$\theta \in [0,2\pi)$}                   
     	\put(88,4){Increasing time}                   
     	\put(90,48){Integrated}                   
        \put(90,46){pixel intensity}
        \put(90,44){(visible)}

        \put(87,30){$\theta = 0, 2\pi$}
        \put(72,38){\color{white}{High-speed}}
        \put(71,36){\color{white}{camera frame}}
        
        \put(31,50){Exhaust flow}
        
        \put(5,44){Combustion products}
		\put(15,43){\vector(1,-1){5.5}}

        \put(0,17){Propellant}
        \put(0,15){regeneration}
        \put(0,13){and mixing}
		\put(4,19){\vector(1,2){2}}

        \put(7,27){Fuel}
        \put(7,25){inj.}
        
        \put(3,37){Oxidizer}
        \put(3,35){inj.}
        
        \put(29,17){Rotating}
        \put(29,15){detonation}
        \put(29,13){wave}
        
        \put(35,22){Oblique shock}
		\put(41,24){\vector(-1,1){5.5}}
        
        \put(18,47){Contact surface}
		\put(26,46){\vector(1,-1){5.5}}
		
        \put(2,50){(a)}
        \put(53,38.5){(b)}
        \put(53,15){(c)}

	    \end{overpic}  
	    \caption{A sketch depicting the canonical flowfield of the Rotating Detonation Engine is shown with the major features annotated. Gaseous fuel and oxidizer is injected through a number of discrete orifices that rapidly mix inside an annular combustion chamber. A supersonic, circumferentially-traveling, reaction wave consumes the newly-mixed propellant, sustaining the motion of the reaction wave. Emanating from this front is an oblique shock wave that sweeps the downstream flowfield. A contact surface marks the location separating the combustion products of this particular reaction front from those of other waves or, in this case, previous round-trips of the same wave. As the reaction front passes over the propellant injectors, the injectors become blocked, as the reaction wave pressure is typically higher than that of injection. A time lag therefore exists before propellant can be re-introduced to the combustion chamber. This time lag is depicted as distance behind the reaction front before the blue reactant zone reappears. Thrust is produced from expelling the hot exhaust products rearward at high velocity and by producing a time- and spatially- averaged high chamber pressure acting on a thrusting wall (in this cartoon, this is the injector face). In the experimental set-up of this study, direct optical access of the annulus allows for the complete space-time history of the detonation waves to be recorded with a high-speed camera. A single frame from an experiment is shown in (b) with the annulus integrated pixel intensity overlaid the annulus location. Stacking each high-speed camera frame gives (c), where line slopes correspond to wave speed. The integrated luminosity trace in (b) corresponds to the trace shown in (c).}
		\label{fig:cutaway}
\end{figure*}
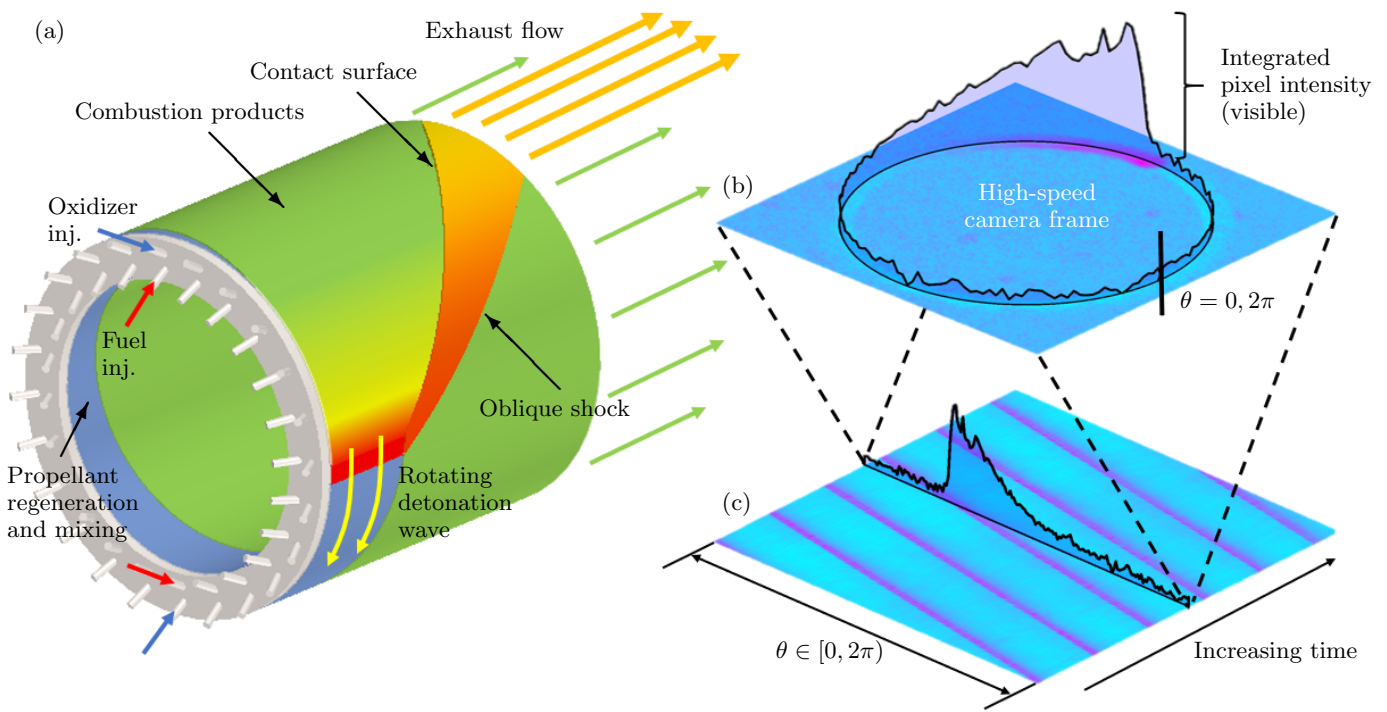

The RDE was first conceptualized as an alternative to standard rocket engine designs in the late 1950’s to early 1960’s  \cite{ssr,NlCHOLLS1966}. In the RDE, instead of the suppression of combustion instabilities, the combustion chamber was designed to leverage and promote a specific and ubiquitous instability which generated rotating combustion fronts whose growth produced a number of discrete co- and counter-propagating traveling detonation waves that consumed injected propellant. The RDE is therefore anomalous in that its steady operation is the saturation of a nonlinear combustion instability, namely the promotion of the self-steepening of the pressure and density gradients caused by heat release in an annular combustion chamber. The collection of detonation waves travel in-sync with the coupled injection and exhaust processes. 

However the RDE is not free from its own set of instabilities: the balance between the nonlinearity of the fluid medium and the competing physical processes of combustion, injection, and exhaust is delicate. Rotating detonation waves have experimentally been found to be very sensitive to combustor boundary conditions (such as inlet pressure and exit plane pressure), propellant heat release, and the geometric parameters of the engine, such as engine length and annulus circumference. For certain conditions, rotating detonation waves have been observed to exhibit a number of remarkable properties that differ significantly from the freely-propagating detonations of conventional theory. A prototypical detonation wave is a \textit{front} that connects the states of unburnt and completely burnt mixture. Rotating detonation waves differ in that they are \textit{pulses}, where the start and end states of the wave are the same. The `tail' of the detonation decays as the burnt gas can expand perpendicularly to the propagation direction of the pulses. Likewise, at a particular point in the annulus, the reactant mixture is regenerated within the transit time of a wave. This balance of heat release (gain), exhaust processes (dissipation), propellant injection (gain recovery), and nonlinearity of the medium governs the pulse shape, number, and behavior \cite{Koch2020}. Should these physical processes be unbalanced, spatially or temporally, a wide array of spatiotemporal dynamics are known to exist and persist, as observed in experiments and detailed computational studies. Such dynamics include mode-locking of pulses \cite{Koch2020}, modulation of the pulse train \cite{Koch2020,Anand2019}, and bifurcations to different numbers of pulses \cite{Koch2020,Bennewitz2019,George2017,Anand2019}. The physical mechanisms and engineering implications of these transients and instabilities are not well understood, especially with regard to operational stability and performance. 

In this article, we investigate the properties of these detonation waves with respect to the interplay of the physical unit processes that govern the pulse behavior. We use a reduced-order mathematical description of the coupling of the physical processes within the combustion chamber to characterize the behavior of the pulses across a wide range of parameter space. We find that the collection of detonation waves exhibit solitonic properties in their interactions and balance physics. However, unlike the solitons of integrable equations, such as the Korteweg–de Vries equation, the solitonic properties of the rotating detonation waves are consequences of the non-local \textit{global} gain dynamics and domain periodicity. Within this context we describe the \textit{fundamental instability} of the rotating detonation wave, namely a Hopf bifurcation from the steadily propagating wave to the time-periodic wave modulation as seen in experiments. Our analysis of the bifurcation structure of the mathematical model shows the universality of this instability: along solution branches of traveling waves, all transients away from the steady case travel through this bifurcation point and develop this instability. We additionally present experimental evidence corroborating these claims. 

In Section \ref{sec:motivation}, we begin by presenting a description of the rotating detonation engine and by reviewing general themes in literature. In Section \ref{sec:observations} we present recent observations of nonlinear dynamics in experiments and in models. In Section \ref{sec:analog}, we present a summary of the RDE analog system and analyze, numerically, the traveling waves admitted by the analog system. In Section \ref{sec:discussion}, we identify and discuss regimes of wave propagation and make the autosoliton analogy. Lastly, we discuss the engineering implications of the present study in section \ref{sec:discussion}.

\section{Background} \label{sec:motivation}
The RDE is an internal combustion engine belonging to a class of engines called \textit{pressure gain combustors}, where the primary mechanism by which heat is added to the flow is through constant-volume combustion. Detonations, or self-sustained supersonic reaction waves where combustion products are sonic relative to the leading shock \citep{Chapman1899,Law2006}, are the naturally occurring physical process that most constant-volume combustors seek to employ, including the RDE. RDEs typically are designed with periodic, annular combustion chambers (Fig. \ref{fig:cutaway}) that provide geometric confinement to the heat release process.  The rapid heat release in the presence this confinement promotes the self-steepening of the gradients of pressure and density within the fluid. Because chemical reactions are accelerated with increasing pressure and temperature, this creates a positive feedback mechanism that further accelerates the reaction front. This front continues to accelerate until the combustion products behind the wave front are exactly sonic relative to the wave, where no `downstream' influences can affect the state of the wave. These waves travel in the circumferential direction ingesting axially moving propellant. After the propellant has been detonation-processed, the hot and high pressure exhaust gasses are ejected through the aft end of the device at high velocity, providing thrust. The path to maturation of RDE technology includes a detailed investigation of the physics of rotating detonation waves, especially the relationship between detonation behavior, engineering performance, and component-level (injection and exhaust hardware, for example) coupling. 

A number of research groups and institutions have successfully sustained rotating detonation waves in annular and disc-type geometries. Although the parameters that uniquely define these specific engines vary drastically across the literature, the behavior of the traveling detonation waves contained within these engines are consistent. Wave speed and count are metrics that are easily observed and readily available in literature. An overarching theme of the RDE community is that the speed and number of waves are related to the energy flux through the engine \citep{Bykovskii2006,Kindracki2011,Wolanski2013,Xie2018,Anand2019}. The wave speeds are decidedly slower than that of the Chapman-Jouguet detonation \citep{Chapman1899}, or the freely-propagating detonation through a premixed flowfield of the same propellant chemistry. Likewise, the wave speeds take on distinct ratcheting transitions when energy flux is taken to be a bifurcation parameter \citep{Bykovskii2006}. In incrementing the number of detonation waves by changing energy (mass) flux, the collection of waves assumes a slower speed than the original state. The opposite scenario also holds, where a decrement in waves results in the remaining waves traveling faster (on the order of 10\% velocity difference \citep{Bykovskii2006,Xie2018,Bennewitz2019,Koch2020}). In addition to transients leading to changes in number of waves, commonly observed are `galloping' detonations, or periodic modulation of detonation wave velocity \citep{Bohon2019,Bennewitz2019,Koch2020}. Such modulation has been identified as a precursor to mode changes \citep{Bennewitz2019}. Amplitude, speed, and phase differences of the waves become time-periodic at the onset of this instability. Such modulation is apparent in the spectral content of fast-response instrumentation, such as in piezo-electric pressure sensors or high-speed camera footage, where spectral sidebands are present and symmetric about a dominant carrier frequency \citep{Boening2018}. Lastly, extreme events such as chaotic propagation, especially during times of ramp-up or ramp-down of propellant feed, have been observed experimentally \citep{Anand2019a}. 

Most laboratory-grade RDEs in literature use gaseous propellants injected through sets of sonic orifices into the combustion chamber. If injection pressure is significantly higher (about a factor of 2) than the combustion chamber pressure, the propellant Mach number becomes one at or near the exit of the injector, meaning the injector is acoustically isolated from the combustion process. No information can be exchanged between the combustion and injection processes: they are decoupled. However, as noted by many (see \citep{Anand2019a}, for example), the detonation waves posses a peak pressure typically an order of magnitude greater than that of the injector feed. The implication is that the acoustic isolation is lost and the injection and combustion processes become coupled. The high pressure detonation waves can induce blockages or backflow into the propellant feed systems. To alter the energy flux through RDEs (fed with gaseous propellant through nominally-choked injectors), one can either change the injector feed pressure or the total injection area. However, these strategies affect the injection-detonation coupling differently \citep{Koch2019,Bach2020}. Consider two sets of injectors delivering equivalent mass fluxes to a combustion chamber: one set with a larger total injection area and lower feed pressure, and one set with smaller total injection area and higher feed pressure. The set with a larger total injection area has a greater potential for coupling with the detonation waves as the detonation peak pressure is much greater than that of injection. Significant coupling of this type generally leads to unstable detonation behavior and/or weakly propagating waves \citep{Koch2019,Bach2020}. Increasing feed pressure such that it becomes comparable to that of the detonation waves decreases the degree of coupling, though using large pressure ratio injectors typically lead to large, unrecoverable pressure loss \citep{Boening2018,Koch2019}. The time scales associated injection and mixing are directly related to these metrics, along with specific injector geometries (the aspect ratio of length to diameter), orientations (axial injection versus radial injection, for example), and mixing schemes (impinging jets or vortical mixing, for example).

Computational Fluid Dynamic (CFD) models of the RDE have been particularly useful in identifying and exploring the physical processes behind the wave behavior and instabilities seen in experiments, including injector coupling. With CFD, one has access to the full state of the system. Therefore, these various wave phenomena can be explicitly linked to performance (combustion efficiency, thrust, etc.) `Unwrapped' two-dimensional (2D) domains of the annular-type RDE have been simulated with great success \citep{Taki1978,Schwer2011,Schwer2011a,Naples2013,Paxson2014,Sousa2017}. These models established the flow field of the RDE with respect to different geometric and operational parameters, and have been instrumental in characterizing the modes and instabilities of rotating detonation waves. In 2D domains, the inlet boundary condition is critical in determining the global behavior of the simulation. A common approach (see \citep{Schwer2011} for a detailed presentation) is to treat the inlet boundary as a solid wall (zero mass flux) if the pressure in the domain is larger than the prescribed injection pressure. If the pressure in the domain is less than the injection pressure, the velocity at the inlet takes on super- or sub-sonic values corresponding to isentropic nozzle expansion of a certain area ratio (injection area relative to combustor annulus area). The prescribed area ratio and injection pressure dictate the time lag behind the passing waves before propellant can be reintroduced to the domain. Long-time history of the propellant regeneration of these simulations show explicitly the `self-adjustment' mechanism \citep{Liu2015} whereby the collection of detonation waves mode-lock to symmetric and maximal phase differences. 

Studies have been extended to three-dimensional (3D) domains \citep{Zhou2013,Frolov2013,Sun2018} to investigate the interaction of traveling detonations on propellant injection, mixing, and exhaust processes. As computing power increases, these full-3D simulations have naturally been refined to match specific hardware and propellants used \citep{Cocks2016,Lietz2019,Pal2020}. Several studies have focused exclusively on the stability and behavior of the traveling waves in 2D and 3D \citep{Yamada2010a,Wu2014,Wu2014a,Yao2015,Fujii2017}. Computational expense becomes a significant problem as one increases complexity in the modeling. Because the flow field of the RDE is inherently low-dimensional (a number of waves traveling at a certain speed is all one needs to recreate the major features of the flow), a number of groups have shifted efforts to simplifying the modeling and analysis to exploit this low-dimensionality \citep{Kawashima2017,Mizener2017,Humble2019} with varying degrees of success. These studies use geometric scaling and `black box' numerical techniques to tie together the physical processes of injection, detonation, and flow expansion and exhaust to determine representative flow fields. With these studies, time-to-solution is prioritized over accuracy. Data-driven methods, like Dynamic Mode Decomposition (DMD), have also been used to extract dominant features from high-speed camera footage and investigate the interaction of these features \citep{Bohon2019a}.

Despite the success of using computational models to determine the RDE flow field and to predict performance with some confidence, the investigation of the fundamental physics governing the behavior of the collection of waves is dilluted by the arbitrarily high-dimensional nature of these formulations. While these studies certainly have merit in the engineering development of specific hardware, they are necessarily constrained to condition- or geometry-specific models and do not generalize to other engines, propellants, boundary conditions, or coupling schemes. These complications drive the state-of-the-art further into increasing fidelity and computational cost. 

In \citep{Koch2020}, we introduced an alternative view of the rotating detonation process. We extended the Majda detonation analog \citep{Majda1981} to model an autowave process; i.e., one that produces solitarily propagating detonative pulses traveling about a periodic domain. The model is a simple 2-component coupled partial differential equation system readily solved by conventional numerical techniques. This model marks a significant departure from the modeling state-of-the-art in that it stresses global energy dynamics and long-time behavior over accuracy and device- and condition-specific computational studies. The approach adequately captures the non-locality (meaning the behavior at a single spatial location is coupled to all other locations) of the energy balance that leads to the diverse behavior seen in experiments. The physics are simplified in this approach: for the pulses to steadily propagate, gain and loss must exactly offset, subject to the nonlinearity (possessing a Burgers'-type flux) and periodicity of the medium. This perspective is adopted from the nonlinear waves community (see \citep{Christov1995,2008a,Purwins2010}). Localized structures that self-organize and propagate as a response to energy pumping from an external source are called dissipative solitons or \textit{autosolitons}.  We classify rotating detonation waves as such. 

The precedent exists for the classification of reaction waves into this mathematical physics framework. Reaction-diffusion systems in active media have been known to exhibit solitonic properties and have a rich mathematical framework with which they can be analyzed \citep{Merzhanov1999}. More topical are flamons \citep{Merzhanov1999} and the phenomenon of spinning reaction fronts \citep{Aldushin1981}. In the engineering community, tangential rocket combustion instabilities have been classified as solitonic 
\citep{Litchford2008}. The pulses of the RDE are differ from those in literature in that the pulse fronts are discontinuous. Our treatment of the reaction front remains rooted in the detonation phenomenon as we retain the nonlinearity that leads to shock formation, whereas the formulations of reaction-diffusion systems neglect this nonlinearity. 

To adopt the view of solitonic propagation of rotating detonation waves has high value. The physical processes responsible for pulse shape and behavior are known. This view therefore de-emphasizes the hardware or condition-specific considerations and does not require the resolution of the complete flow field. Potentially the greatest benefit of this perspective are the ability to determine relationships between stability, performance, and model parameters.

Note that in this article we have restricted our discussion to co-rotating waves exclusively, despite the prevalence of counter-rotating in literature. The inclusion of counter-rotating waves is the subject of future studies.

\begin{figure*}[]
        \centering
        \begin{overpic}[width=1.0\linewidth]{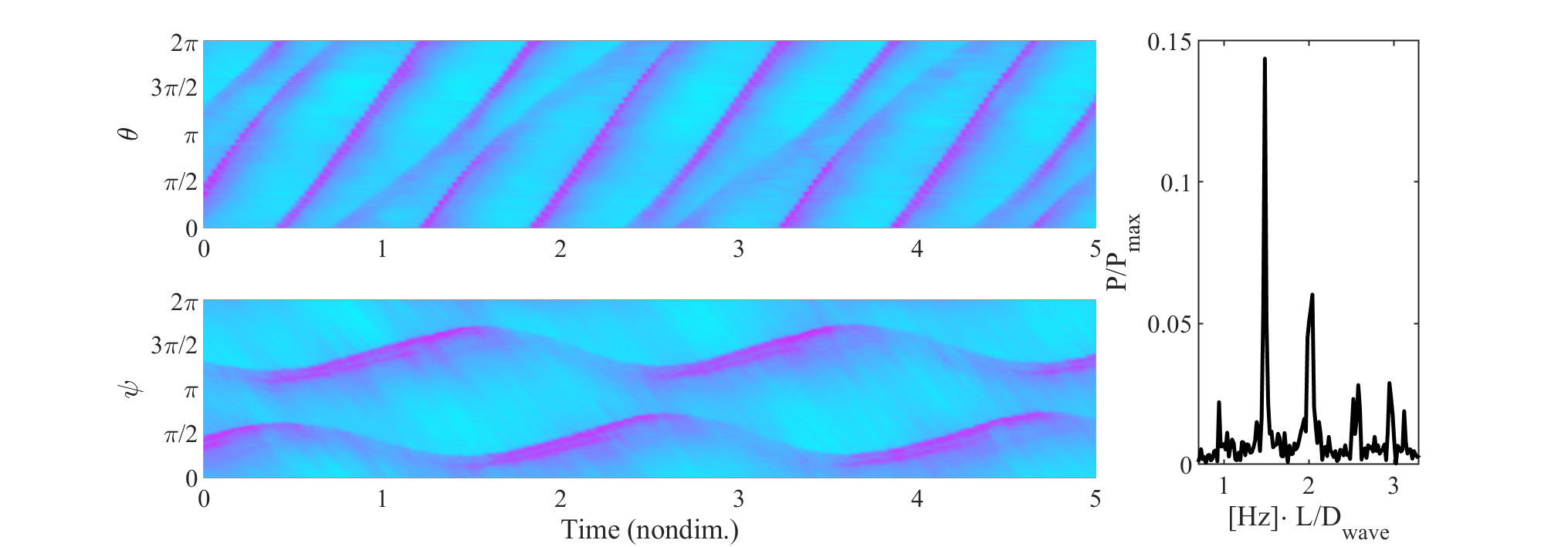}
        \put(13.5,30){(a)}
        \put(13.5,13.5){(b)}
        \put(77,30){(c)}
	    \end{overpic}  
	    \caption{Raw integrated pixel intensity of RDE annulus through time is shown in (a) for an experiment with large-amplitude modulation. The two waves present in the domain exchange strength and amplitude in a regular fashion. At each wave collision, the waves nonlinearly interact, leaving an observable phase shift in the trajectories of the waves. The data shown in (a) is recast into the mean-velocity reference frame in (c). Here, the oscillations in phase difference between the waves is explicit. The accompanying spectrum in (c) shows the frequency content in terms of wave count (unit of abscissa). Sidebands exist symmetric about the wave count frequency of two, though dominating the frequency content is the lower sideband as this experiment is near a bifurcation point to one wave. }
		\label{fig:2WaveModulate}
\end{figure*}

\begin{figure*}[]
        \centering
        \begin{overpic}[width=1.0\linewidth]{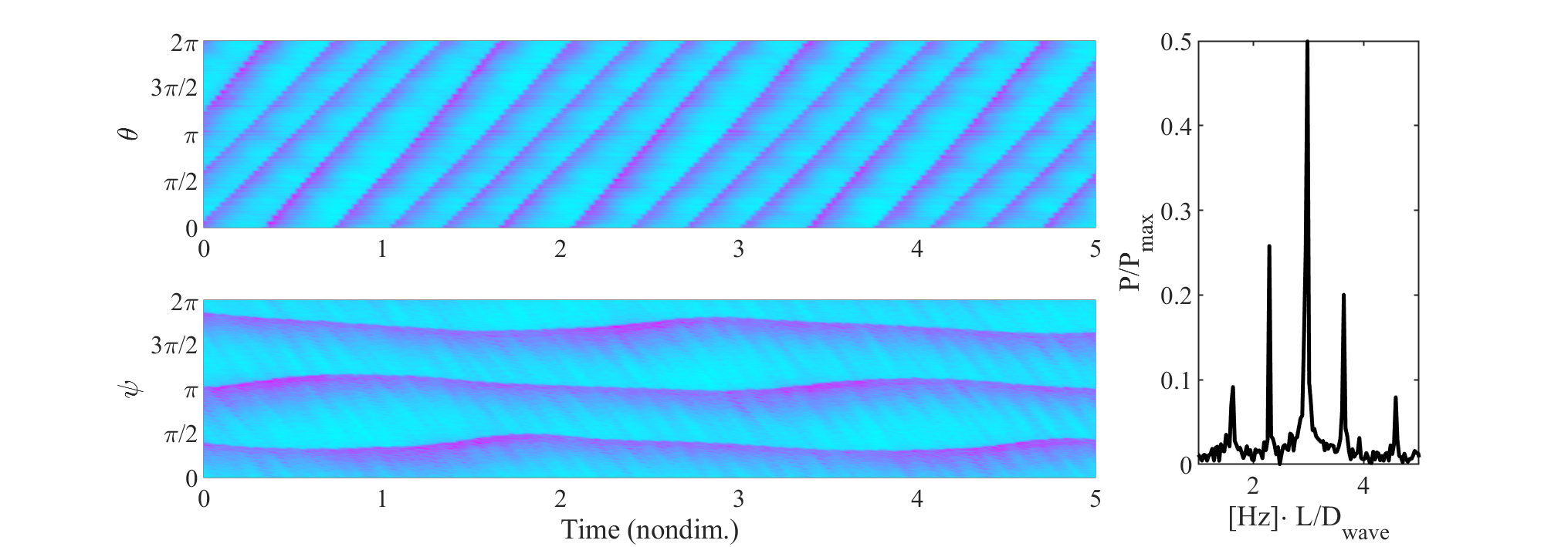}    
        \put(13.5,30){(a)}
        \put(13.5,13.5){(b)}
        \put(77,30){(c)}
	    \end{overpic}  
	    \caption{Shown in (a) is raw pixel luminosity for a three-wave modulation case in the laboratory reference frame. Recasting this to the mean-velocity reference frame (b), the oscillations in phase difference become explicit. Although this instability is the same type as displayed in the two wave case of Fig. \ref{fig:2WaveModulate}, the amplitude of the modulation is less severe. The spectrum of the experiment is shown in (c), with sidebands near wave counts of two and four symmetric about the carrier frequency of three waves. }
		\label{fig:3WaveModulate}
\end{figure*}

\section{Experiments} \label{sec:observations}
For this article, we have collected a series of observations from experiments that highlight the spatio-temporal dynamics of the detonation waves. The rotating detonation engine used is a gaseous oxygen- and methane-fed engine based on a 76mm outer annulus diameter and 76mm core length. Each experiment consists of a minimum of 0.5 seconds of `hot' operation where the feed lines and combustion chamber pressures have settled to steady values. The engine is mounted to a large dump volume that captures all engine exhaust. Because the system is closed, both the inlet (feed pressure) and outlet (dump volume pressure) boundary conditions can be controlled and independently set. Likewise, routing the exhaust into a dump volume has allowed for the safe installation of an optical viewport roughly 2 meters downstream of the engine. Using a high speed camera, the complete spatio-temporal history of the detonation waves is recorded for each experiment. These space-time histories are the primary metric by which we compare our model to experiments. A single frame from an example high-speed recording is shown in Fig. \ref{fig:cutaway}. The location of the annulus is overlaid with black circles. By integrating the pixel intensity around the annulus for each frame of the video \citep{Bennewitz2019a}, a profile of the luminosity can be constructed and stacked through time to yield a spatio-temporal history (Fig. \ref{fig:cutaway}) of the detonation waves in an experiment. In this view, slopes correspond to wave angular velocity. For the experiment shown in Fig. \ref{fig:cutaway}, the wave motion is steady through time.

\begin{figure*}[]
        \centering
        \begin{overpic}[width=1.0\linewidth]{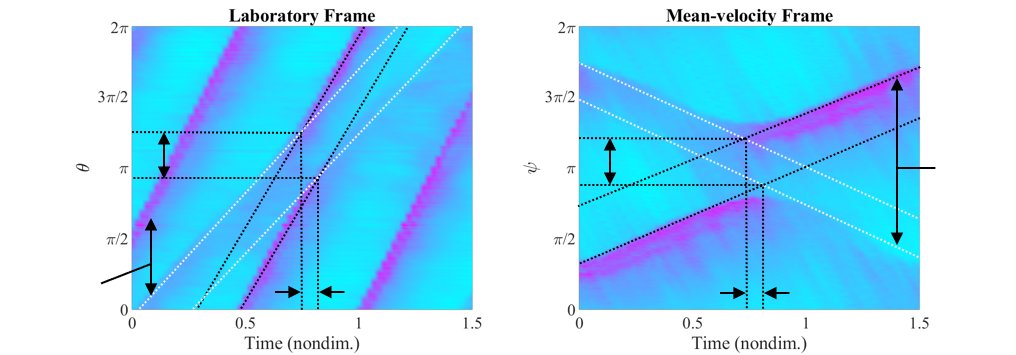}   
        \put(3.5,7){Phase}
        \put(3.5,5){difference}
        
        \put(92.5,19){Phase}
        \put(92.5,17){difference}
        
        \put(32,7){$\Delta t$}
        \put(76,7){$\Delta t$}

        \put(13,19){$\Delta \theta$}
        \put(57,18.5){$\Delta \theta$}
        
        \put(13.5,30){(a)}
        \put(57.5,30){(b)}
        
	    \end{overpic}  
	    \caption{A single period of modulation for the data in Fig. \ref{fig:2WaveModulate} in the laboratory reference frame is shown in (a). The two waves present in the combustion chamber interact nonlinearly, producing observable time and phase shifts. Such kinematic traces are hallmarks of solitonic interactions. The data in (a) is shown in the mean-velocity reference frame in (b). }
		\label{fig:solitons}
\end{figure*}

We present here a subset of experiments that exhibit a modulational instability with the goal of characterizing this phenomena with a reduced-order model in Section \ref{sec:analog}. Figures \ref{fig:2WaveModulate} and \ref{fig:3WaveModulate} show different numbers of waves in the annulus that travel with a modulated speed, amplitude, and phase difference. For these figures, we plot these histories against time normalized by the average round-trip time of a wave. The time units correspond to number of round trips. Raw integrated pixel luminosity for a two-wave modulation case is given in Fig. \ref{fig:2WaveModulate}a. In viewing this space-time history, it is apparent that in this experiment the waves have two modes of propagation. The first mode is characterized by a stronger (higher amplitude), faster moving pulse. The second mode is characterized by a significantly weaker, slowly moving pulse. The two co-existing waves in the RDE annulus regularly alternate between these propagation modes. The point at which the waves `switch' modes is at the local maxima of the modulation; i.e., when the two waves are closest together. At this close range, a fundamentally different balance physics exists that changes the behavior of the waves. The strengths and speeds of the waves are directly tied to the amount and distribution of available reactant in the annulus. Immediately after a wave-pair interaction, the (now) faster of the two waves has an excess of reactant through which it can propagate stably (with a constant velocity and amplitude). This imbalance of propellant distribution exists because: (i) the weaker wave has not blocked propellant injection to the degree that the stronger wave has, and (ii) a temporal imbalance exists corresponding to the large amplitude phase differences of the waves. Because the stronger wave travels faster than its counterpart, it approaches the tail of the slower wave, where the amount of renewed reactant is significantly less than required to sustain the speed and amplitude of the strong wave. This strong wave therefore decelerates, as the dissipative processes (exhaust) now dominate the physics. At this point of interaction, the phase difference between the strong and weak waves is small - on the order of $45$ degrees or $\pi/4$ radians. The accompanying phase difference - the one preceding the weak wave - is therefore $2\pi - \pi/4$. With these large phase differences (and accompanying time lags), reactant regeneration asymptotically approaches a state of complete `refill', where no combustion products are present in the flow and the reactant is fully mixed. The weaker wave now rapidly gains strength: the input energy to the wave overwhelms the dissipative processe. Finally, the transition to the stronger mode of propagation is complete when the energy input to the wave exactly balances the dissipative processes. This saturation of the growth of the waves is explicitly seen in the spatio-temporal history of the experiment: the paths of the waves are straight lines (paths of constant velocity) connected by brief periods of nonlinear wave-to-wave interaction. Displayed in Fig. \ref{fig:3WaveModulate}a is similar time-periodic modulation, though with three waves. 

All experiments possessing this modulational instability share a similar spectrum. A carrier frequency corresponding to the average velocity (or, if normalized as in Figs. \ref{fig:2WaveModulate}c and \ref{fig:3WaveModulate}c, a count of the waves in the domain) is accompanied by sidebands symmetric about the carrier frequency. By recasting the wave trajectories into the reference frame of the velocity corresponding to the carrier frequency, one can visualize the same dynamics as deviations from the average velocity. The sidebands in the spectra correspond to frequencies near that which would appear for an increment or decrement in number of waves. For example, for the three wave case of Fig. \ref{fig:3WaveModulate}, the spectrum shows a dominant (carrier) frequency of three waves with sidebands at approximately two and four waves. Figure \ref{fig:2WaveModulate}b is the representation of the data contained in Fig . \ref{fig:2WaveModulate}a recast into the reference frame of the mean velocity of the waves by the transformation $\psi = \theta -ct$, where $c$ corresponds to the speed associated with the carrier frequency of the spectrum. In this reference frame, the oscillations of phase difference between the waves is explicit, as is the modulation of wave amplitude. The lower-amplitude modulations of Fig. \ref{fig:3WaveModulate} do exhibit the same characteristics of the two wave case, through the range of interaction for these cases is observably larger than that of the two wave case. 

In the brief wave-to-wave interactions, the waves exhibit solitonic behavior. The strong wave assumes the shape and velocity of that of the weaker and is displaced by a small phase shift. Likewise, the weak wave assumes the shape and velocity of that of the stronger wave - again displaced by a small phase shift. This interaction is most easily observed with close-scale interactions. In Fig. \ref{fig:solitons}, displayed is a single period of oscillation extracted from the two wave modulation case of Fig. \ref{fig:2WaveModulate}. The time shifts $\Delta t$ yielding phase shifts $\Delta \theta$ give the interaction the appearance of a solitonic collision.

\section{The RDE Analog System} \label{sec:analog}
Our goal is to use a reduced-order mathematical formulation to (i) reproduce, qualitatively, the modulational instability and solitonic interactions of collections of rotating detonation waves, and (ii) characterize the conditions under which these instabilities develop. We first summarize the mathematical model first presented in Koch et al. \citep{Koch2020}. This model builds upon the Majda detonation analog to account for dissipation, reactant regeneration, and periodic boundaries. Using this model, we run a sweep of numerical simulations to survey wave behavior. Lastly we use numerical continuation to extract the traveling wave branches (and deviating branches) as a function of a bifurcation parameter. 

\subsection{Model Formulation}
The model quantifies the spatio-temporal evolution of a property analogous to specific internal energy, $u(x,t)$, on a one dimensional (1D) periodic domain:
\begin{equation} \label{eq:u}
\frac{\partial u}{\partial t} +u\frac{\partial u}{\partial x} = q\left( 1 - \lambda \right)\omega(u)  - \epsilon u^2
\end{equation}
\begin{equation} \label{eq:lambda}
\frac{\partial \lambda}{\partial t} = \left( 1 - \lambda \right)\omega(u) - \lambda\beta(u),
\end{equation}
where $q$ is the propellant heat release, $\epsilon$ is a loss coefficient, $\omega$ is the rate law for the combustion kinetics, and $\beta$ is the injection model.  The model is normalized such that $x$ maps to $\theta\in[0,2\pi)$ with periodic boundary conditions imposed due to the circular structure of the rocket engine (See Fig. 1).

In this study, we treat the chemical kinetics as Arrhenius in type, with the functional form of:
\begin{equation}
\omega(u) = k \exp \left( {\frac{u-u_c}{\alpha}} \right),
\end{equation}
where $k$ is a pre-exponential factor, $u_c$ is prescribed `ignition temperature', and $\alpha$ is analogous to activation energy. Note that as written, the kinetics are auto-catalytic: $\omega(u) > 0$ always. The evolution of the combustion progress variable $\lambda$ is governed by the competition of combustion (following the rate law of $\omega(u)$) and injection (following the injection model $\beta(u)$). For injection, we use an activation function that sufficiently mimics injection from choked orifices:
\begin{equation}
\beta(u) = \frac{su_p}{1 + \exp \left({r(u - u_p)} \right)},
\end{equation}
where $su_p$ is the time constant for the regeneration of $\lambda$. In real engines, $s$ is influenced by the injection scheme, mixing effectiveness, and total injection area. The quantity $u_p$ corresponds to injection pressure. This form asymptotically assumes the value of the numerator if the value of $u$ is small compared to $u_p$. In the limit as $u$ becomes much larger than $u_p$, $\beta(u)$ approaches zero. Similarly, if either $s$ or $u_p$ is zero, there is no regeneration of $\lambda$. In this manner, the parameter $u_p$ is used to control the injection sensitivity threshold. Lastly, the quadratic loss term, $\epsilon u^2$, is the imposition of expansion processes on the exhaust-side of the traveling waves.

The model's steady (or quasi-steady) state exists when the energy flux into and out of the domain balance. This condition is met when the integrated losses are equal to the integrated heat release over the domain:

\begin{equation} \label{eq:flux}
\dot{E}_{domain} = \int_0^L q(1 - \lambda)\omega(u) dx - \int_0^L \epsilon u^2 dx 
\end{equation}

For steady planar deflagration fronts and mode-locked traveling wave solutions, the integrals exactly balance and $\dot{E}_{domain} = 0$. For oscillatory plane waves, $\dot{E}_{domain} \neq 0$: there is periodic accumulation and ejection of energy in the domain that oscillates in-phase with injection and heat release, similar to the autoignition model of Frank-Kamenetskii \citep{frank,frank1955}.

Transient phenomena, such as the initial start-up of a simulation or immediately after ignition of an RDE, exhibit an imbalance of the integrals in Eqn. \ref{eq:flux}: $\dot{E}_{domain} > 0$. From the initial condition of zero combustion, the losses in the chamber are minimal. An accumulation of $u$ will occur until the domain satisfies the relationship in Eqn. \ref{eq:flux}. A direct consequence of this behavior is the influence of the accumulation of $u$ on the kinetic model. As $u$ increases with the onset of combustion, the chemical reactions governed by the simplified Arrhenius kinetics are \textit{accelerated}. This feedback mechanism promotes deflagration in the entirety of the domain, which can then transition to a number of detonations. This process is the physical mechanism for wave nucleation in RDEs.

\subsection{Modulation of the Deflagration State}
\begin{figure*}[]
        \centering
        \begin{overpic}[width=1.0\linewidth]{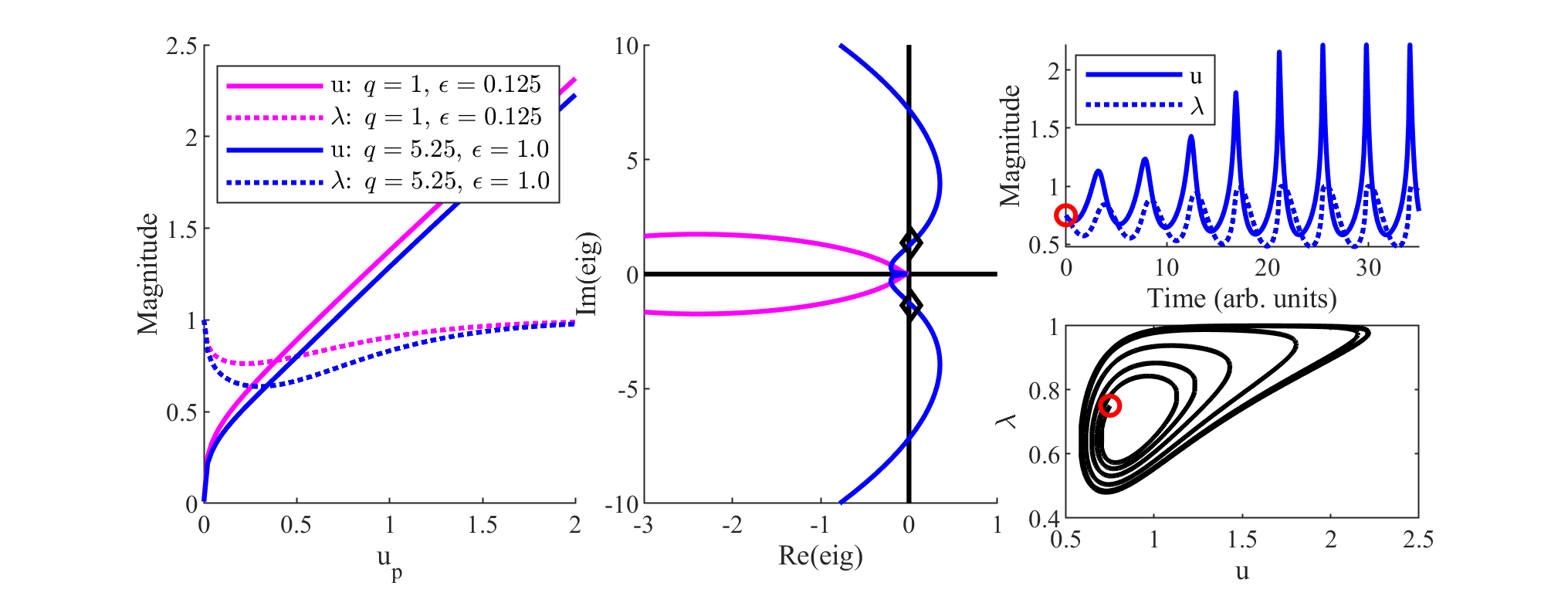} 
        \put(13.3,35){(a)}
        \put(41.5,35){(b)}
        \put(68.5,35){(c)}
        \put(68.5,15.5){(d)}
     	\put(75,35){$u_p=0.55$}
	    \end{overpic}  
	    \caption{(a) Fixed points of the deflagration system of Eqs. \ref{eq:deflagration1} and \ref{eq:deflagration2} for the parameters listed in Table \ref{tab:plainsimulations}. The spectrum of the linearized system evaluated at the fixed points is shown in (b). A Hopf bifurcation to a stable limit cycle exists at $u_p \approx 0.55$. The eigenvalues of the black diamonds in (b) correspond to those of the system simulated in (c) and its associated phase plane in (d). The initial condition for the simulation is the red circle in (c) and (d). }
		\label{fig:deflagration}
\end{figure*}
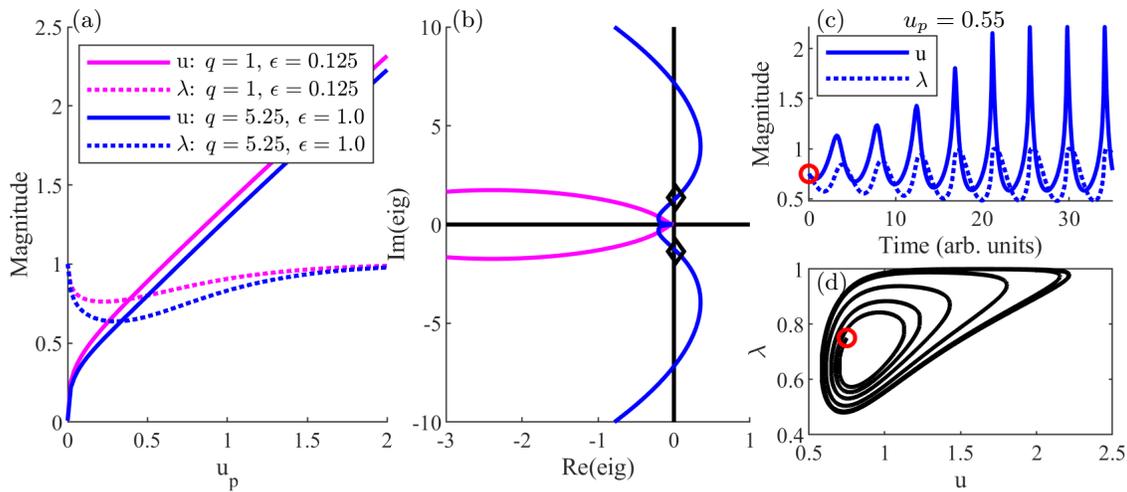

We first consider the plane wave case. 
Modulational instabilities play a critical role in driving the overall dynamics of a similar damp-driven system of mode-locked lasers~\cite{Koch2020}, where the instability/stability of plane wave solutions determine the overall global dynamics~\cite{agrawal2012nonlinear,bronski1996modulational}.
Plane waves satisfy the further-reduced coupled ordinary differential equations:
\begin{equation} \label{eq:deflagration1}
\frac{du}{dt} = qk\left(1-\lambda\right)\exp{\left(\frac{u-u_c}{\alpha}\right)} - \epsilon u^2
\end{equation}
\begin{equation} \label{eq:deflagration2}
\frac{d\lambda}{dt} = k\left(1-\lambda\right)\exp{\left(\frac{u-u_c}{\alpha}\right)} - \frac{s u_p \lambda}{1 + \exp{\left(r \left(u - u_p\right)\right)}}
\end{equation}
Stationary solutions (a steady, planar deflagration front) exist for $\epsilon u^2 =  q\left(1-\lambda\right)\omega(u)$ and $\left(1-\lambda\right)\omega(u) = \lambda\beta(u)$. These fixed points can be found numerically. Figure \ref{fig:deflagration}a displays fixed points for the deflagration system for the parameters listed in Table \ref{tab:plainsimulations}. The spectrum of the linearized system evaluated at the fixed points yields Fig. \ref{fig:deflagration}b. For small dissipation and heat release, the steady planar deflagration is linearly stable. For high energy flux conditions, an instability can form leading to the growth of oscillations and a stable limit cycle (as shown in Fig. \ref{fig:deflagration}b-d). This is a Hopf bifurcation with the injection sensitivity threshold ($u_p$) as the bifurcation parameter. This instability physically corresponds to the axial injector-combustion-exhaust resonance experimentally observed by many in literature \citep{Anand2019a,Koch2020}.

\begin{table} 
	\caption{Plane Wave Simulation Parameters}
	\label{tab:plainsimulations}
	\centering
	\begin{tabular}{ccccc}
	\hline
	$\alpha$ &$u_c$ & $s$ & $k$ &$r$\\
	
	0.3 & 1.1 & 1.0 & 1 & 5\\
	\hline
	\end{tabular}
\end{table}

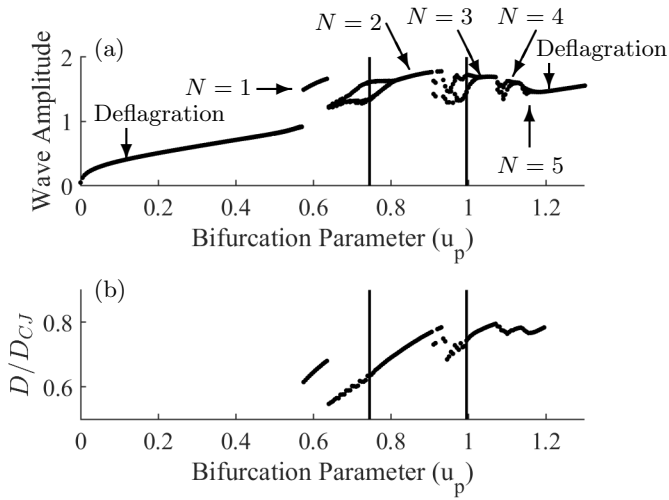
\begin{figure}[]
        \centering
        \begin{overpic}[width=1.0\linewidth]{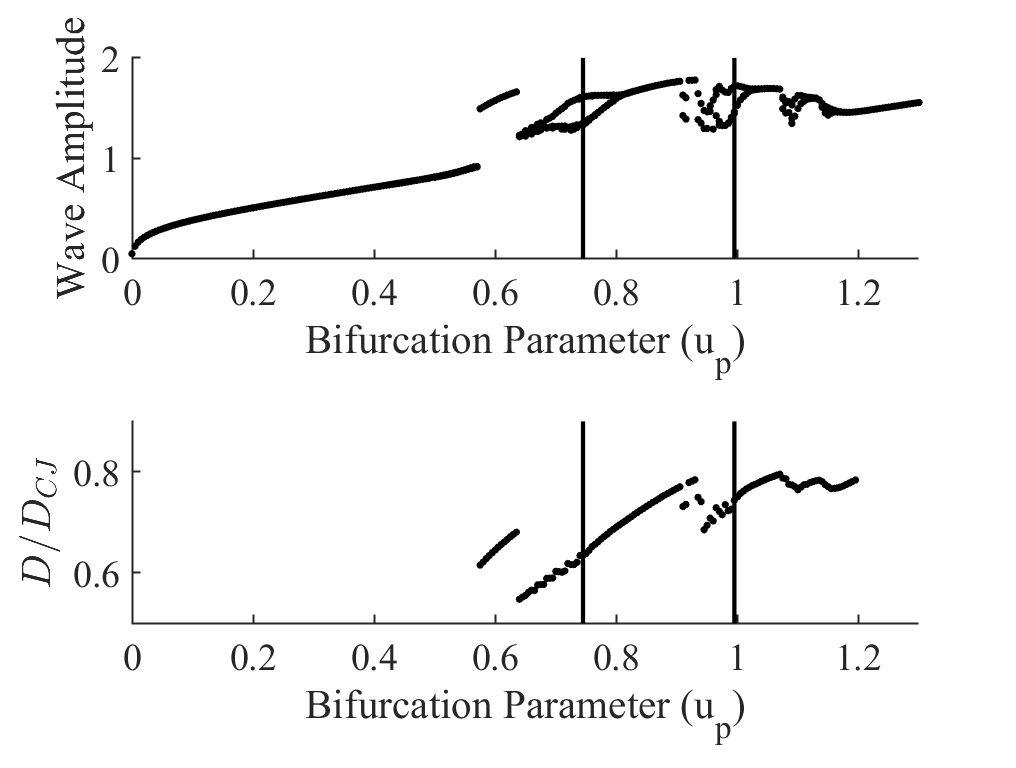}  
     	\thicklines
		\put(40,65){\vector(1,0){5.5}}
     	\put(29,64){$N=1$}
       	
 		\put(60,75){\vector(1,-2){3.5}}
     	\put(49,74){$N=2$}    	   
     	
		\put(71,74){\vector(1,-2){3.5}}
     	\put(63,75){$N=3$}    
     	
		\put(83,74){\vector(-1,-2){3.5}}
     	\put(77,75){$N=4$}   

		\put(82,56){\vector(0,1){6.5}}
     	\put(77,52){$N=5$}  
     	
    	\put(15,60){Deflagration} 
		\put(20,59){\vector(0,-1){5}} 

    	\put(83,70){Deflagration} 
		\put(85,69){\vector(0,-1){4}} 
		
		\put(15,70){(a)}
        \put(15,33){(b)}
     	
	    \end{overpic}  
	    \caption{A bifurcation diagram of the RDE model analog showing peak amplitude of the simulated domain is shown in (a). Model parameters are listed in Table \ref{tab:simulations}. In increasing the bifurcation parameter $u_p$ from zero, the system initially exhibits planar deflagration fronts, then traveling waves (from 1 to 5 waves), then back to a deflagration front. The associated speeds of the traveling waves are given in (b). }.
		\label{fig:dnsBif}
\end{figure}

\subsection{Bifurcation Structure of Traveling Waves}
We use the open-source finite volume code \textit{PyClaw} \citep{Ketcheson2012} to perform the direct numerical simulations of the model and the Matlab-based software \textit{pde2path} \citep{pde2path,Uecker2019} to perform the numerical continuation. However, the model system as written in Eqns. \ref{eq:u} and \ref{eq:lambda} admits solutions with discontinuities. Although \textit{PyClaw} is well-suited to handle shocks, \textit{pde2path} was originally intended for systems of elliptic partial differential equations. To facilitate the bifurcation analysis, the model system is necessarily regularized with diffusion such that the solutions become continuous, albeit still possessing sharp gradients characteristic of the reaction fronts. We therefore perform our bifurcation study of the modified system:
\begin{equation} \label{eq:cU}
\frac{\partial u}{\partial t}  = \nu_1 \frac{\partial^2 u}{\partial x^2} -u\frac{\partial u}{\partial x} + kq\left(1-\lambda\right)\exp{\left(\frac{u- u_c}{\alpha}\right)}  - \epsilon u^2
\end{equation}
\begin{multline} \label{eq:cL}
\frac{\partial \lambda}{\partial t} = \nu_2 \frac{\partial^2 \lambda}{\partial x^2} + k\left(1-\lambda\right)\exp{\left(\frac{u - u_c}{\alpha}\right)} \\ - \frac{su_p\lambda}{1 + \exp{\left(r\left( u - u_p \right)\right)}},
\end{multline}
where the constants $\nu_1$ and $\nu_2$ are diffusivities associated with the combustion (diffusing $u$) and injection (diffusing $\lambda$) processes, respectively. The model parameters used in this study are listed in Table \ref{tab:simulations}.

\begin{table} 
	\caption{Traveling Wave Simulation Parameters}
	\label{tab:simulations}
	\centering
	\begin{tabular}{ccccccccccc}
	\hline
	$L$& $q_0$ &$\alpha$ &$u_c$ & $s$ & $k$& $\epsilon$ &$r$& $\nu_1$ & $\nu_2$ &$D_{CJ}$\\
	
	$2\pi$ & 1.0 & 0.3 & 1.1 & 3.5 & 1 & 0.15 & 5 & 0.0075 & 0.0075 & 2\\
	\hline
	\end{tabular}
\end{table}

A bifurcation diagram showing the peak amplitude of the domain as a function of the parameter $u_p$ (injection sensitivity threshold) is shown in Fig. \ref{fig:dnsBif}, as computed by numerical simulation on a converged grid with PyClaw. Each simulation was initialized with a singe localized pulse $u(x,0) = (3/2){sech}^{20}(x-1)$ with a 'half combustion' condition of $\lambda(x,0) = 0.5$.

At $u_p = 0$, the injection term $\beta$ is zero and the entire domain dissipates to a zero value. As $u_p$ increases, a planar deflagration front forms: the dissipation term ($-\epsilon u^2$) first dominates the dynamics of the domain, prohibiting the formation of stable pulses, then relaxing to exactly balance the input energy given by a non-zero $\beta$. At a critical value of $u_p \approx 0.56$, the initial pulse can form a single stably-propagating wave of a significant amplitude. At this condition, the input-output energy balance is still satisfied, but the time scale corresponding to the round-trip time of the wave (the speed of which is determined by the energy release associated with the Arrhenius kinetics) has become comparable to the time scales of gain regeneration ($su_p$) and dissipation. Continuing to increase $u_p$ increases the peak amplitude of the single wave until $u_p \approx 0.65$, where a transition to two waves occurs. This transition marks the point at which the mean value of $u$ in the domain has accelerated the kinetics to the point where the effects of parasitic deflagration (combustion that is not associated with the traveling waves) and detonative combustion on the domain are of the same order. The single wave's amplitude decreases as the parasitic deflagration is consuming a increasing amount of the available energy. Once the parasitic deflagration can self-steepen to form a shock during the round-trip time of a detonation wave, a deflagration-to-detonation transition occurs and the number of waves increases by one. This transition is seen in Fig. \ref{fig:dnsBif} when increasing $u_p$ above 0.65 (a transition from one to two waves) and again around 0.9, 1.07, and 1.1. As $u_p$ becomes large (beyond $u_p \approx 1.1$), the domain regresses back to a planar deflagration front. The time scale of the kinetics is now much faster than all others in the model, including that of the traveling waves. All input energy input is quickly consumed and dissipated.

\begin{figure}[]
        \centering
        \begin{overpic}[width=1.0\linewidth]{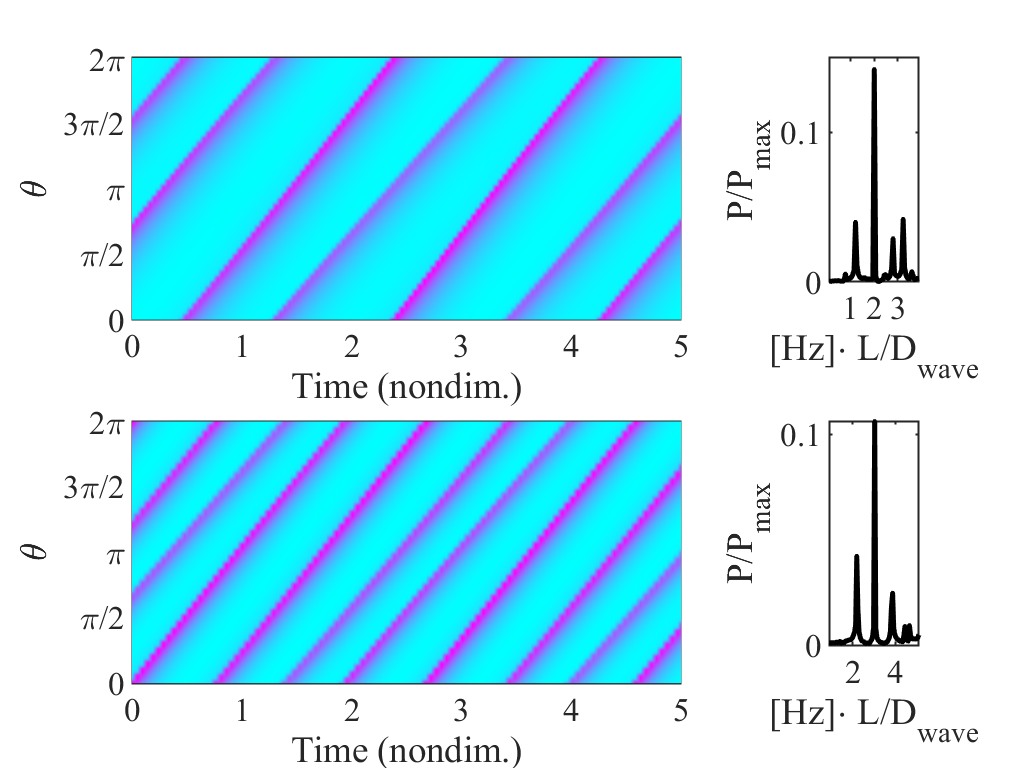}  
	    \end{overpic}  
	    \caption{Modulated wave trajectories in the laboratory reference frame from numerical simulations corresponding to the vertical lines in Fig. \ref{fig:dnsBif}.}
		\label{fig:simModulate}
\end{figure}

The inter-pulse regions of the bifurcation diagram show a diverse set of behavior, including wave modulation. In Fig. \ref{fig:dnsBif}, the vertical lines correspond to the simulation histories of Fig. \ref{fig:simModulate} for two and three wave cases. For this region of operability space in $u_p$, the waves travel unsteadily with modulation similar to that which is observed in experiment (Figs. \ref{fig:2WaveModulate} and \ref{fig:3WaveModulate}). In the bifurcation diagram of Fig. \ref{fig:dnsBif}, the wave modulation for the two wave branch is bounded by stable two wave propagation on both sides of the instability. The modulation region for the three wave branch is bounded by a jump to the two wave branch and by stable three wave propagation. 

The bifurcation diagram of Fig. \ref{fig:dnsBif} shows structure. A planar delfagration branch exists and is predominantly linear with $u_p$. Branches exist for each number of traveling waves that interact in some manner, giving intervals of $u_p$ where the wave dynamics are not steady. Transitioning from steady propagation (a number of waves moving at constant velocity) to unsteady propagation constitutes a bifurcation to an instability.

Using the numerical simulations from Fig. \ref{fig:dnsBif} as initialization seeds, we construct an estimate of the complete bifurcation diagram using pde2path for the parameters listed in Table \ref{tab:simulations}. The system of Eqns. \ref{eq:cU} and \ref{eq:cL} are modified to produce steady profiles by including an imposed offsetting advection velocity and performing the 2-parameter continuation with $u_p$ and the imposed velocity. See Appendix \ref{app:hopf} for the complete formulation of the continuation problem. 

\begin{figure}[]
        \centering
        \begin{overpic}[width=1.0\linewidth]{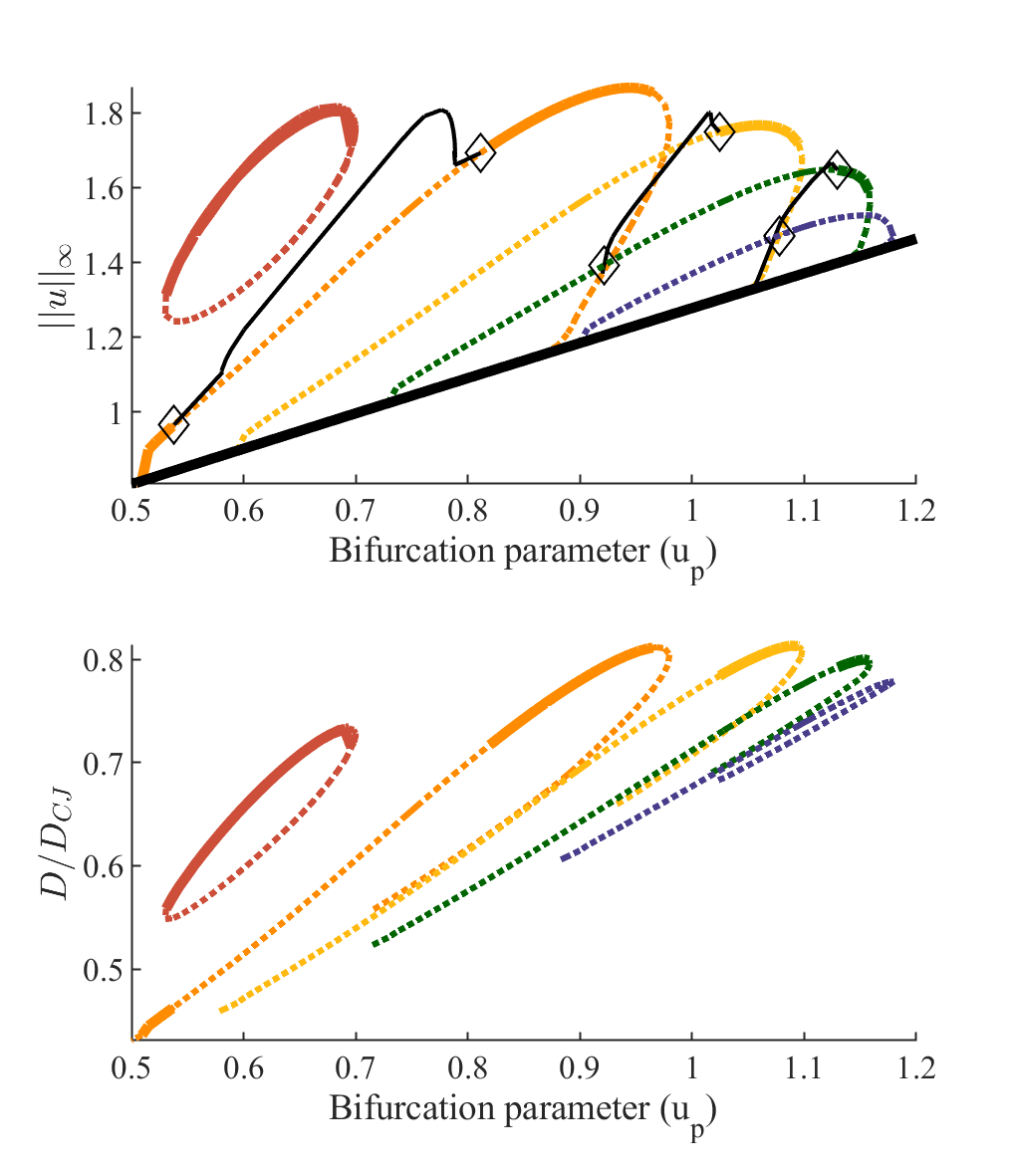}  
     	\put(14,90){$N=1$}
       	
     	\put(44,94){$N=2$}    	   
     	
     	\put(62,90){$N=3$}    
     	
     	\put(73,86){$N=4$}   

     	\put(77,81){$N=5$}  
     	
    	\put(37,60){Deflagration Branch} 
    	\thicklines
		\put(50,63){\vector(0,1){7}} 
		
		\put(14,95){(a)}
        \put(14,47){(b)}

	    \end{overpic}  
	    \caption{Bifurcation diagram replicating that of Fig. \ref{fig:dnsBif} computed from numerical continuation. Emanating from the trivial deflagration branch are 5 traveling wave branches. For each traveling wave branch, solid lines correspond to stable propagation and bashed lines correspond to unstable solutions. The single-wave traveling branch is a closed ring of solutions - an isola. Corresponding wave speeds along the traveling wave branches are shown in (b). Hopf bifurcations exist at the transition from stability to instability for the traveling wave branches, marked by diamonds. Along the Hopf branches are time-periodic modulations of wave speed, amplitude, and phase difference.}
		\label{fig:pde2path}
\end{figure}

Figure \ref{fig:pde2path} contains several distinct branches of solutions for the model system: one `trivial' branch and five traveling wave branches. Along these branches, solid lines indicate regions of stability whereas dotted lines indicate unstable regions. The `trivial' branch is the locus of points satisfying the input-output energy balance with no contributions from traveling waves. This is the \textit{deflagration} branch: the locus of solutions where a planar front spanning the domain consumes and quickly dissipates all input energy. Because of the viscous regularization of the model system, there exists a small region of stability around this branch where diffusion inhibits wave growth. We note that in the system without viscous regularization, this is not necessarily the case. Any change in concavity with Burgers' type flux (without diffusion) leads to wave growth and shock formation, and therefore a perturbation off the deflagration branch may indeed lead to wave formation if the local energy gain exceeds the local energy dissipation.

The single traveling wave branch is a closed solution branch - an isola - that exhibits stability for the top half of the branch. The region of stability is bounded by fold bifurcations at the extremes of the isola. The solution branches of higher number of waves are qualitatively similar to the single-wave branch: each possesses a region of stability (with the exception of the five wave branch, which is everywhere unstable for these model parameters) bounded by bifurcations to instability.

The wave speeds along the branches vary dramatically - by a factor of two for some branches - though they saturate at about the same value across branches. Relative to the CJ speed of detonations for the Majda detonation analog ($D_{CJ} = 2q = 2$), their speed is about 80 - 90\% of the theoretical maximum.  As $u_p$ and the number of waves increases, there is marked drop in wave speed and amplitude, consistent with direct numerical simulations and experiments with large wave counts \citep{Anand2019}. 

\subsection{Self-Similarity and Domain Length}
\begin{figure}[]
        \centering
        \begin{overpic}[width=1.0\linewidth]{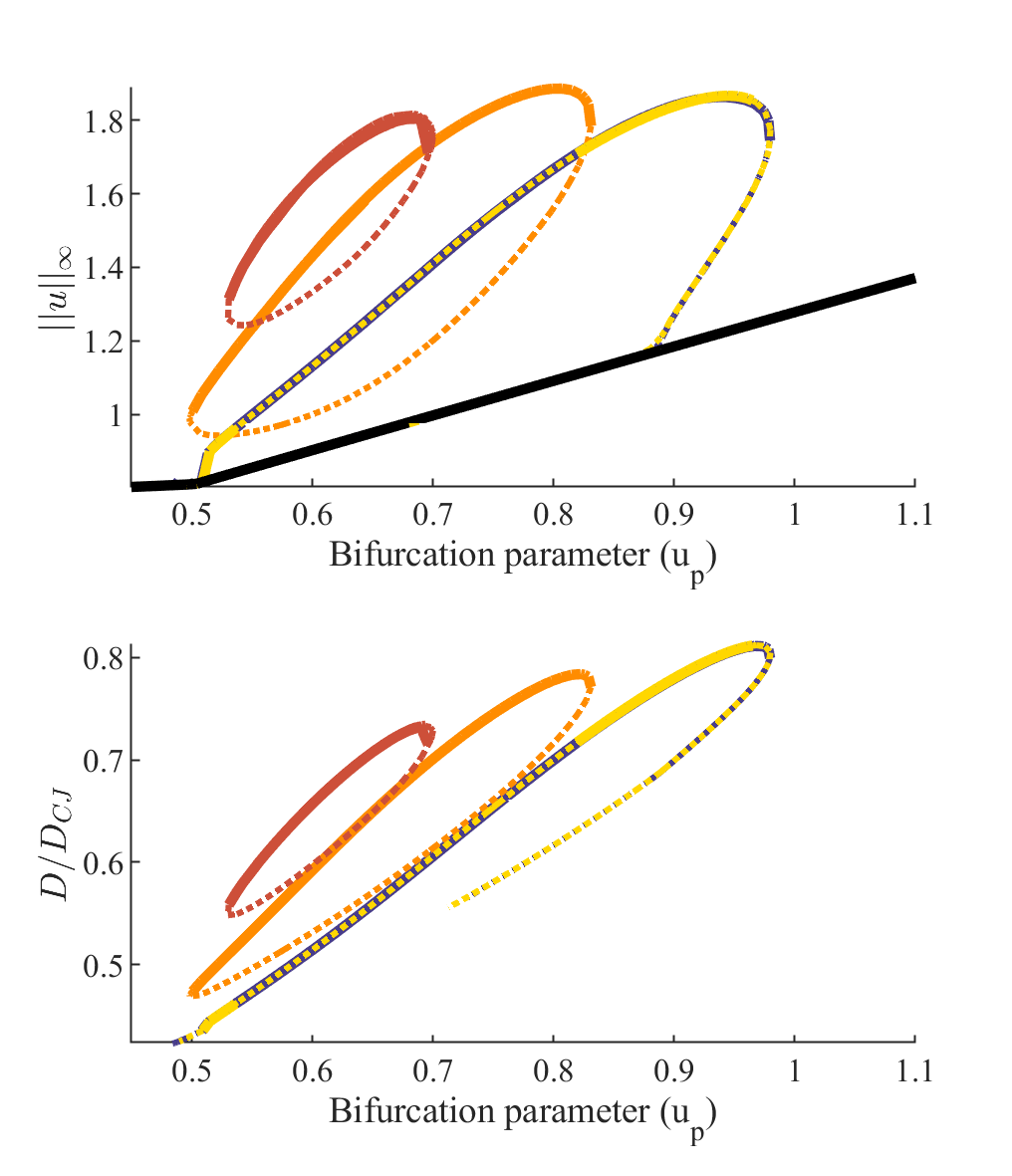}  
        
     	\put(14,91){$N=1,L=2\pi$}
     	\put(36,94){$N=1,L=3\pi/2$}     	

     	\put(68,86){$N=1,L=\pi$}
     	\put(75,83){and}
    	\put(68,80){$N=2,L=2\pi$}
     	
    	\put(37,60){Deflagration Branch} 
    	\thicklines
		\put(50,63){\vector(0,1){5}}         
        
		\put(14,95){(a)}
        \put(14,47){(b)}
        
	    \end{overpic}  
	    \caption{(a) A bifurcation diagram showing $N=1$ and $N=2$ branches of the original model system (domain length of $L=2\pi$) alongside two additional cases: $N=1$ on $L=3\pi/2$, and $N=1$ on $L = \pi$. Note that the traveling wave branches are self-similar - the curves of $N=1$ on $L=\pi$ and $N=2$ on $L=2\pi$ overlay identically, including the wave speeds, shown in (b). The maximum wave speed and amplitudes for the model system with parameters listed in Table \ref{tab:simulations} shows strong dependence on the domain length, in effect changing the time scale for the round-trip time of the detonation wave.}
		\label{fig:pde2pathScale}
\end{figure}
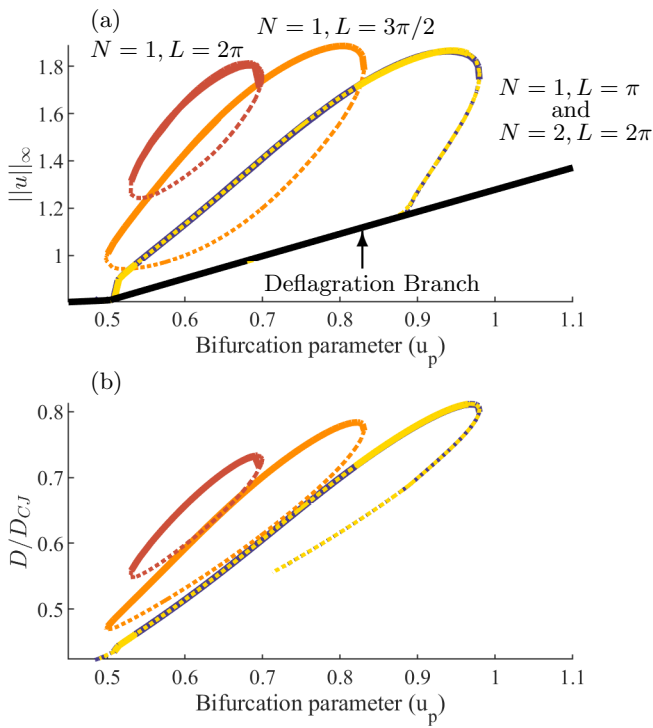

In Fig. \ref{fig:pde2pathScale}, the single- and double-wave branches of Fig. \ref{fig:pde2path} plotted alongside branches of the same system but different domain lengths: $L=2\pi$ (the original system), $L = 3\pi/2$, and $L=\pi$. The traveling wave branch for $N=1$ on a $L=\pi$ domain is identical to that of the $N=2$ wave branch on a $L=2\pi$ domain. The system is self-similar, scaled by domain length. Furthermore, the wave speeds are along these branches are also equivalent. Increasing the domain size from $L=\pi$, the $N=1$ branch detaches from the deflagration branch and forms an isola. With further increase in domain size, the isola decreases in size. A point of criticality exists where the domain is too large to support a single wave: the transit time of the wave becomes too long compared to the time required for parasitic deflagration to self-steepen and form an additional wave. At this point of criticality, the $N=1$ isola ceases to exist, though the $N=2$ branch detaches from the deflagration branch and forms a new isola. This process of isola formation and destruction repeats indefinitely with each doubling of the domain length. Note, however, that the regions of stability of the self-similar branches ($N=1$ on $L=\pi$ and $N=2$ on $L = 2\pi$) are not consistent: for this set of parameters, instability only exists for wave count greater than one.

\subsection{The Hopf Bifurcation to Wave Modulation}
\begin{figure*}[]
        \centering
        \begin{overpic}[width=1.0\linewidth]{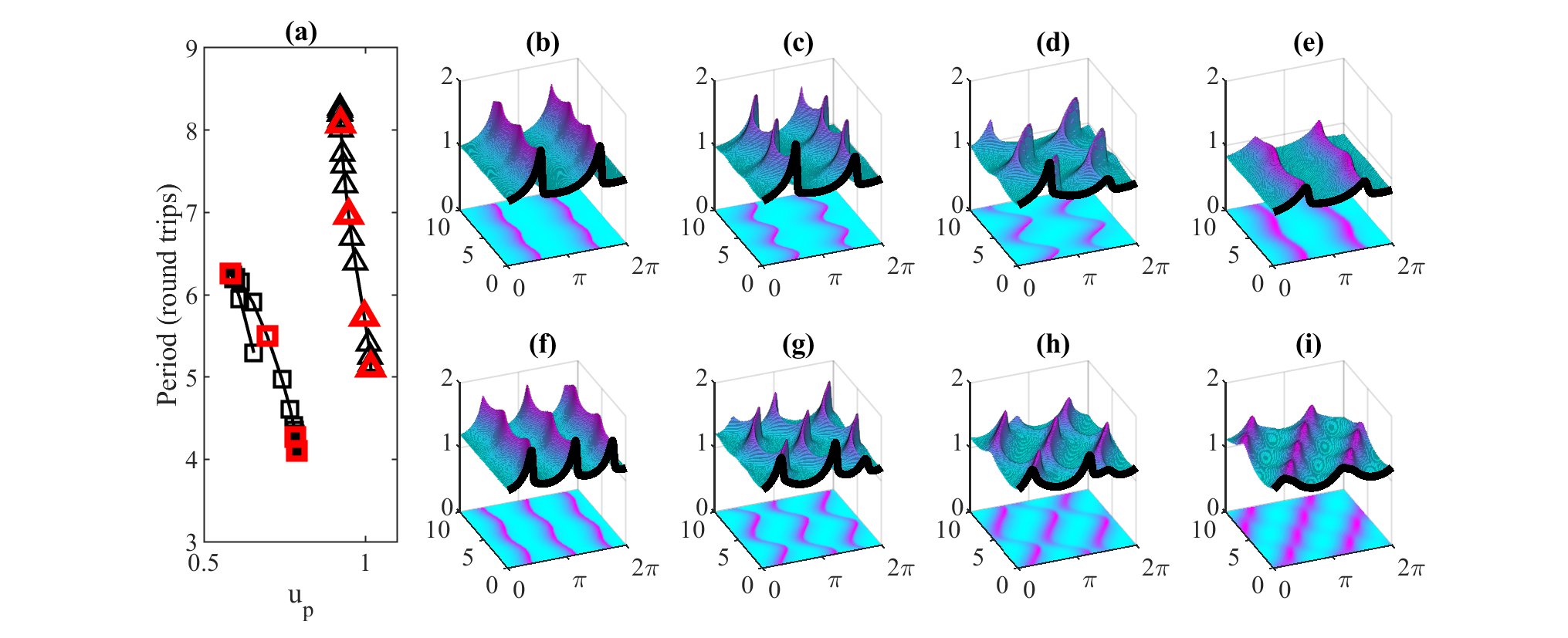} 
       	\thicklines
        
		\put(16,10){\vector(1,1){2}}
     	\put(14,9){(b)}
       	
 		\put(15.5,16){\vector(1,-1){2}}
     	\put(13.5,16){(c)}    	   
     	
		\put(19,24){\vector(-1,-2){2}}
     	\put(18,25){(d)}    
     	
		\put(14.5,27){\vector(0,-1){3}}
     	\put(13.5,28){(e)} 
     	    	
		\put(23.5,11){\vector(0,1){5}}
     	\put(22.5,9){(f)}   

		\put(20.5,19){\vector(1,1){2}}
     	\put(18.5,17.5){(g)}          

		\put(19,28){\vector(3,-1){2}}
     	\put(16.5,28){(h)}          
 
		\put(18,33.5){\vector(4,-1){3}}
     	\put(16,33){(i)}

	    \end{overpic}  
	    \caption{The two- and three-wave branches of the bifurcation diagram of Fig. \ref{fig:pde2path} each posses regions of stability and instability. Hopf bifurcations from the steadily traveling wave branches exist at these transitions and are reproduced in (a). By continuing the Hopf branches, one can extract a diverse set of potential modulational behavior. In (a), the period of the orbits along the two- and three-wave Hopf branches are displayed by number of wave round trips (based on wave speed averaged over one period) as a function of the bifurcation parameter. Example Hopf orbits for the two wave branch are shown in (b)-(e) and in (f)-(i) for the three wave branch. These orbits are displayed in the mean-velocity reference frame. Along the Hopf branches, the amplitude of the phase differences and the waves vary dramatically. For the three wave branch, in the extreme limit of each wave-pair interaction, the stronger of the two interacting waves overruns the weaker wave resulting in a reduction of number of waves by one. This phenomenon is shown in (i).}
		\label{fig:orbits}
\end{figure*}

The traveling wave branches of Fig. \ref{fig:pde2path} each posses a region of stability: the detonations propagate stably with a constant velocity through time. However, there exist Hopf bifurcations that spawn branches of periodic orbits away from the stably-propagating pulse train. By adding an appropriate Hopf constraint (see Appendix \ref{app:hopf}), the continuous branch of orbits - the Hopf branches - can be extracted. For the two and three wave cases, these branches are shown in Fig. \ref{fig:orbits} with example solution plots. Note that for these branches, we have not evaluated stability - we have only traced the branch location in parameter space.

The branches intersect the traveling wave branches at two points. At each intersection is a Hopf bifurcation. Along these Hopf branches, the period of oscillation and the amplitudes of phase differences, wave amplitudes, and wave speeds are all modulated: the branches constitute the possible states of modulation for the given parameters.  These variations in propagation behavior are exhibited in Fig. \ref{fig:orbits}. At each intersection with the traveling wave branches (the end points of the curves in Fig. \ref{fig:orbits}a), the modulation is low in phase difference amplitude, though the modulation onset frequency is about a factor of two faster for the higher energy (larger $u_p$) cases. 

At each local extreme of phase difference oscillations, the waves interact solitonically as in the experiments of Figs. \ref{fig:2WaveModulate} and \ref{fig:3WaveModulate}. From the collection of waves, the pair that interact exchange strength and undergo a phase shift. This is clearest in Figs.  \ref{fig:orbits}d and \ref{fig:orbits}g-\ref{fig:orbits}h. At the onset of modulation (approached from a high $u_p$), the phase differences between the interacting wave pairs is large, as are the apparent phase shifts. As the oscillations in phase differences grow, the phase shifts between interacting pairs decreases. In the extreme limit of the three wave Hopf branch, the phase shifts become such that the weaker of the interacting waves is overrun and the phase shifts are zero, resulting in the reduction of the number of traveling waves. Similar phenomena occur for Hopf branches of higher wave count, as shown in Fig. \ref{fig:pde2path} along the 4-wave Hopf branch.

\begin{figure*}[]
        \centering
        \begin{overpic}[width=1.0\linewidth]{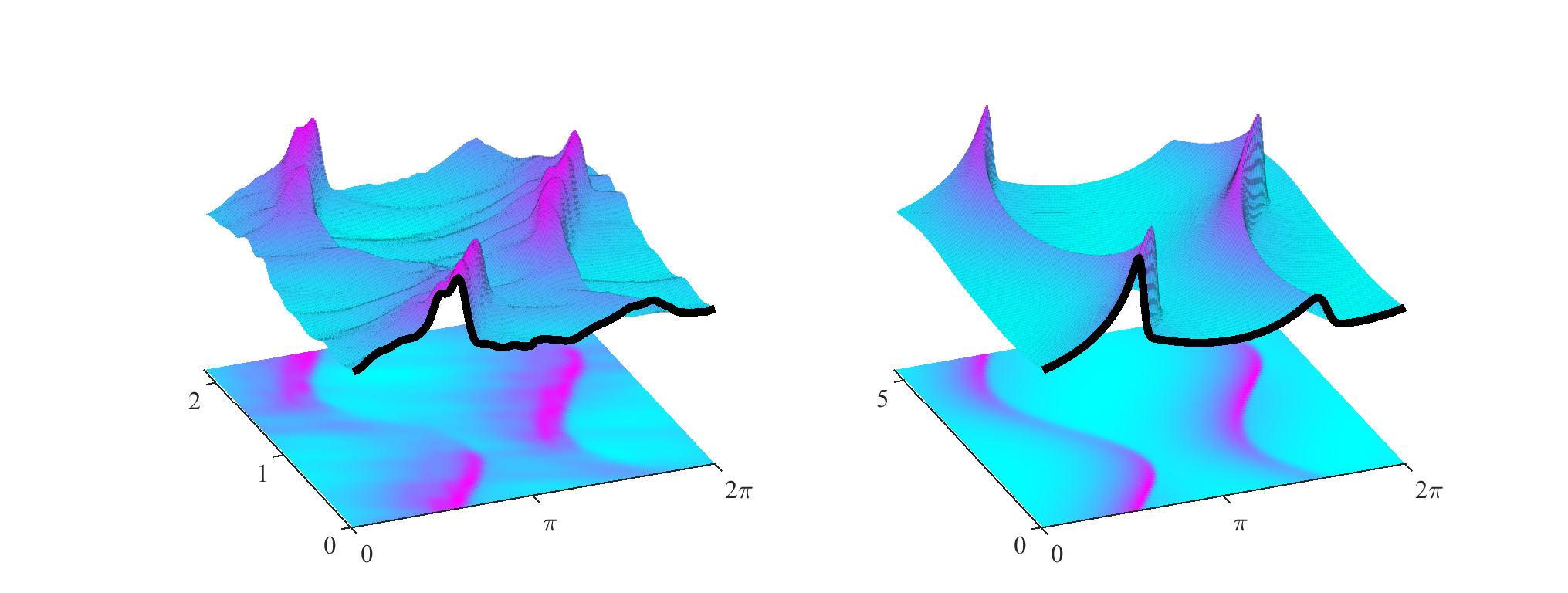} 
        \put(13,6){$T$}    
        \put(36,2){$\psi$}         
        \put(58,6){$T$}    
        \put(80,2){$\psi$}   
        
      	\put(24,32){Experiment} 
    	\put(70,32){Model} 
            
	    \end{overpic}  
	    \caption{Extracted Hopf orbits after the onset of a modulational instability in an experiment and in the numerical continuation of the RDE analog system. The two detonation waves interact through global gain dynamics, producing a distinct and repeatable kinematic trace. The model is in good qualitative agreement with the experimentally obtained orbit.}
		\label{fig:modulationCompare}
\end{figure*}
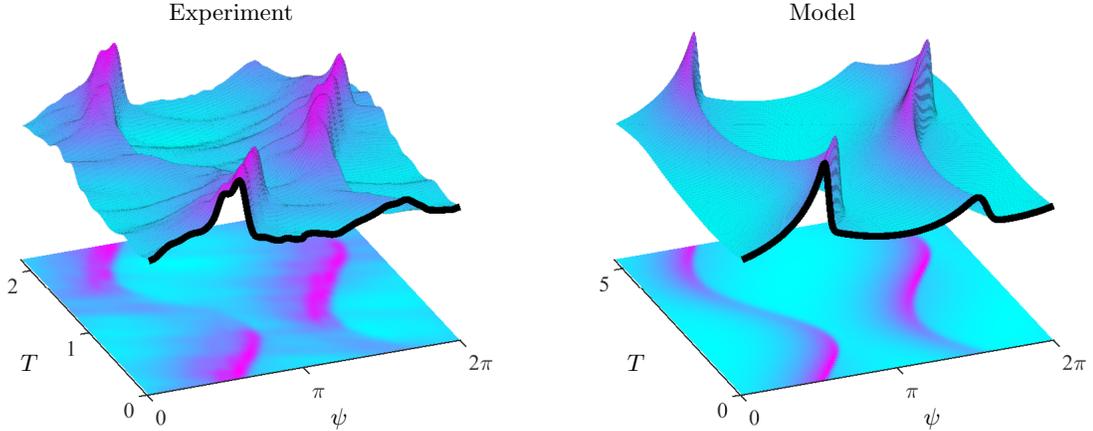

Figure \ref{fig:modulationCompare} compares a single-period of oscillation for two rotating detonation waves in an experiment and in simulation of the analog system. Both are displayed in the average-speed reference frame. The behavior of the waves in the experiment and model share the same qualitative behavior, including exchange of wave strength, and similar phase shifts through the interaction. However, the period of oscillation (shown in number of wave round-trips around the domain, based on average wave speed) of the model does not match that of the experiment, though the model likely is not quantitatively matching all timescales of interest (chemical, injection, mixing, and wave transit time). Note that we have not performed any parameter selection techniques nor have we fit the model to data. Provided the aforementioned time scales are such that the model can support traveling detonation waves, the bifurcation structure and wave interactions are qualitatively similar. 

\section{Discussion} \label{sec:discussion}

Rotating detonation waves have been experimentally observed to exhibit solitonic propagation, especially in wave pair interactions, where wave strengths are swapped and phase shifts are imposed (Fig. \ref{fig:modulationCompare}). In performing numerical simulations and a numerical bifurcation analysis of the rotating detonation analog system, we qualitatively reproduced the behavior seen in experiments. Thus, the rotating detonation wave analog sufficiently mimics real-engine behavior, including the bifurcation structure, wave selection transients, and modulation of detonation waves.

In this section, we wish to emphasize the main findings of this study and how they relate to the development of the rotating detonation engine. First, we establish that the global gain dynamics are the mechanisms responsible for the observed physics. Next, we discuss the scales of interaction in the system and the solitonic behaviors they produce. Within this context, we establish the modulational instability of rotating detonation waves as the \textit{fundamental instability} of the system. Lastly, the engineering implications of this study are discussed, including stability, control, and scaling.

\subsection{Global Gain Dynamics}
The process of mode-locking of rotating detonation waves implies a significant communication pathway between the waves. This is a direct contradiction to classical detonation theory. Steady detonations are supersonically moving fronts, meaning they are `unaware' of the fluid ahead of the wave. Similarly, the combustion products behind the detonation wave travel away from the shock front at a velocity sonic relative to the wave front. This implies that there are no characteristics that can propagate from the burnt side of the detonation upstream to the location of heat release. The rotating detonation engine possesses two physical constraints on the problem that lifts the restriction of wave (communicative) isolation. First, the domain is periodic, not infinite or pseudo-infinite in many classical studies, and second, the reactant ahead of the detonation waves is a function of the cumulative history of the detonation waves. 

Periodicity means that the waves see the tail of the preceding wave (or, in the case of one wave, its own tail). The behavior of the detonation waves is necessarily dependent not only on the local combustion at the shock front (where classical detonation theory stops), but also the time scales for energy dissipation and propellant recovery. These three physical processes have four different time scales corresponding to: (i) combustion, (ii) round-trip time of the detonation wave, (iii) dissipation (exhaust processes), and (iv) gain recovery (injection and mixing). These time scales vary by several orders of magnitude in real engines, with combustion being the fastest (sub-microseconds) and dissipation being the slowest (millisecond).

These time scales are related. The time scale of combustion is related to the amount of reactant ahead of the comubstion zone and the quality of the mixing processes. The transit time of the wave is governed by the heat release associated with the detonation wave (Chapman-Jouguet theory), but also by the properties of the ingested gas, as they are not necessarily at standard or injected conditions. Likewise, the properties of the fluid in the inter-pulse space is governed by the slower-scale dynamics of the exhaust processes. The rotating detonation analog succeeds in capturing the dynamics seen in experiments because the physical processes and their time scales are included and properly coupled.

We therefore claim that the \textit{global gain dynamics} are responsible for the observed dynamics. It is insufficient to consider exclusively the physics of the front - classical detonation theory in this case - to describe the dynamics of the collection of waves. This approach also contrasts the convention of the field of autosolitons. Typical solitary structures in driven-dissipative systems are held together because of the \textit{local} balance of gain, loss, nonlinearity, and dispersion. In the RDE, the local balance physics cannot be decoupled from the global gain dynamics.

\subsection{The Fundamental Instability}
The modulation seen in RDE experiments and simulations has been characterized as `galloping' rotating detonation \citep{rdestability,Wolanski2013,Bennewitz2019}. This term is adopted from the phenomenon whereby one dimensional detonations undergo a Hopf bifurcation to front modulation followed by period-doubling to chaotic propagation \citep{Ng2005,Henrick2006}. In these studies, activation energy is typically taken to be the bifurcation parameter. Qualitatively, the behavior of `galloping' detonation is very similar to the modulation seen in RDEs: the frontal motion of the waves and peak pressures oscillate through time, just as in Figs. \ref{fig:2WaveModulate} and \ref{fig:3WaveModulate}, though naturally the studies of `galloping' detonation are performed on pseudo-infinite, 1-dimensional domains.

The viscous Majda detonation analog \citep{Majda1981} does not exhibit the aforementioned nonlinear dynamics of real detonations. The \textit{nonlinear stability} of detonations of the viscous Majda detonation analog has been evaluated with rigor with Evans function-based techniques \citep{Lyng2004,Jung2012,Humpherys2013} - indeed, no Hopf bifurcation exists "in a normal parameter range or in the limit of high activation energy" \citep{Humpherys2013}. As the RDE analog system is fundamentally a modification of Majda's model to include global gain dynamics, we claim that the inclusion of these terms and the restriction to a periodic domain is responsible for the formation of the Hopf bifurcation to modulation. Furthermore, the physical mechanisms responsible for the modulation in these cases are fundamentally different: for rotating detonation waves, this is the interplay of the processes contributing to the global gain dynamics. In `galloping' detonations, the mechanism is encapsulated solely within the local frontal dynamics, including those of induction, reaction, and expansion of gases. By extension, we claim that this type of modulational instability found in RDEs is \textit{unique} and \textit{fundamental} to the RDE. As seen from the bifurcation diagrams of Figs. \ref{fig:pde2path} and \ref{fig:pde2pathScale},  for a given number of waves, to arrive at any behavior other than steady propagation, the operating point \textit{must} first pass through this Hopf bifurcation. 

\subsection{Engineering Implications}
\subsubsection{Stability and Control}
In this study, we established for the first time stability limits on rotating detonation waves and have begun to characterize the instabilities present in the system. Knowing the operability maps of the engine (analogous to the bifurcation diagrams of Fig. \ref{fig:pde2path}) and regions of instability means that an engine controller and actuation schemes can be conceptualized to maneuver about the operability maps, especially in the presence of multi-stability and hysteresis. This generalizes for multi-parameter operability maps and bifurcation diagrams. Additionally, sensing can be performed by a small number of point-measurements of the frequency spectrum (via fast-response piezoelectonic pressure sensors, for example). The operating frequency spectrum, domain geometry, and wave count fully define the location of the operating point in the bifurcation diagrams.

\subsubsection{Performance and Design}
Although the RDE analog system is still qualitative in nature, we can nevertheless examine metrics related to performance. In experiments, a readily observed metric is wave speed. In classical detonation theory, detonation wave speed is directly related to the heat release associated with the wave \citep{Chapman1899}. However, in the RDE, this is a misleading metric for performance. Because the detonation physics cannot be decoupled from the global gain dynamics, the measured wave speed is a property of the \textit{system}, not only of the detonation wave front. To exemplify this, we refer to experimental studies \citep{Koch2019} where decidedly shallow-fronted pulses exist in RDE chambers traveling at speeds comparable to the acoustic velocity of combustion products. However, if related to the Chapman-Jouguet speed of detonation for standard conditions, these speeds are of the same order. The physical difference in these scenarios is the mean combustor state. In the RDE, the properties of the fluid ingested by the wave are significantly higher temperature and pressure than a 1-dimensional detonation propagating through standard conditions. If the upstream state of a detonation wave increases in temperature and pressure (or in the RDE analog system, an increase in $u$), \textit{less} heat release is required to attain the Chapman-Jouguet condition for detonation formation \citep{Law2006}. Wave speed as a performance metric is therefore misleading without quantitative measurements of the fluid prior to detonation wave arrival. 

An example of the interplay between wave speed and base state of the fluid in the domain is shown in Fig. \ref{fig:pde2pathScale}. Among the different domain lengths, the peak wave speed and wave amplitudes do not coincide. In general, the fastest waves have lower base-to-peak amplitudes than that of the local maximum. Because the RDE analog system also captures parasitic deflagration, we propose an alternative metric by which one can compare the state of the system to the `trivial' deflagration branches of Figs. \ref{fig:dnsBif}, \ref{fig:pde2path}, and \ref{fig:pde2pathScale}. This is simply the average of the square of the domain: $\overline{u^2}$. The analogous physical property is combustion chamber pressure \citep{Koch2020}.

We postulate that to form strongest detonation waves, the time scales of injection, exhaust, and the transit time of the waves should match. Significant attention has been devoted to injection and mixing within these engines, effectively shortening their associated time scales. However, decreasing this time scale \textit{exclusively} will only lead to an increased affinity for parasitic deflagration. As an example, we refer to the premixed rotating detonation engine experiments from the United States Air Force Research Laboratory \citep{Andrus2016}. In this experimental study, during operation of the engine, stationary flames were attached to the injection sites in an extreme version of parasitic deflagration. The traveling waves propagated at a speed comparable to the acoustic speed of the combustion products of the propellant. In this scenario, it is probable that the time scales associated with combustion were far too short to match the relatively long time scale of dissipation. Perhaps the most straightforward manner in which the time scale of the exhaust processes can be shortened to match those of injection and mixing is by lifting the constraint of constant-annular area ducts. By allowing the flow to expand through an area expansion, the flow can travel, supersonically, away from the combustion zone. This is not necessarily the case for constant-area annular ducts, where a strong pressure gradient is required to push the flow away from the combustion zone towards a mechanical or thermal choke point, increasing the residence time of a fluid particle as it travels through the RDE \citep{Fotia2016,Koch2019a}.

\begin{figure*}[]
        \centering
        \begin{overpic}[width=1\linewidth]{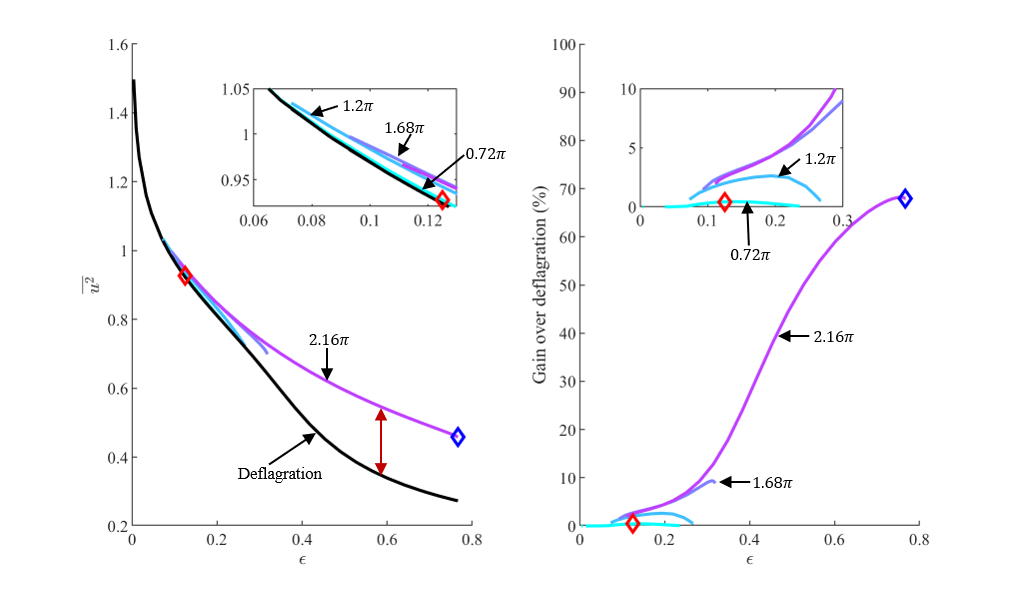} 
           
	    \end{overpic}  
	    \caption{In (a),  $\overline{u^2}$ is displayed for both the deflagration and single-wave solution branches for varying domain lengths as a function of the dissipation coefficient. The percent improvement over the deflagration branch (taken as the magnitude of the red arrow in (a) normalized by the value of the deflagration branch at that location) is shown in (b). In the limit of weakly dissipative rotating detonation mode, the traveling wave branches merge with the deflagration branch.}
		\label{fig:performance}
\end{figure*}

\begin{figure}[]
        \centering
        \begin{overpic}[width=1\linewidth]{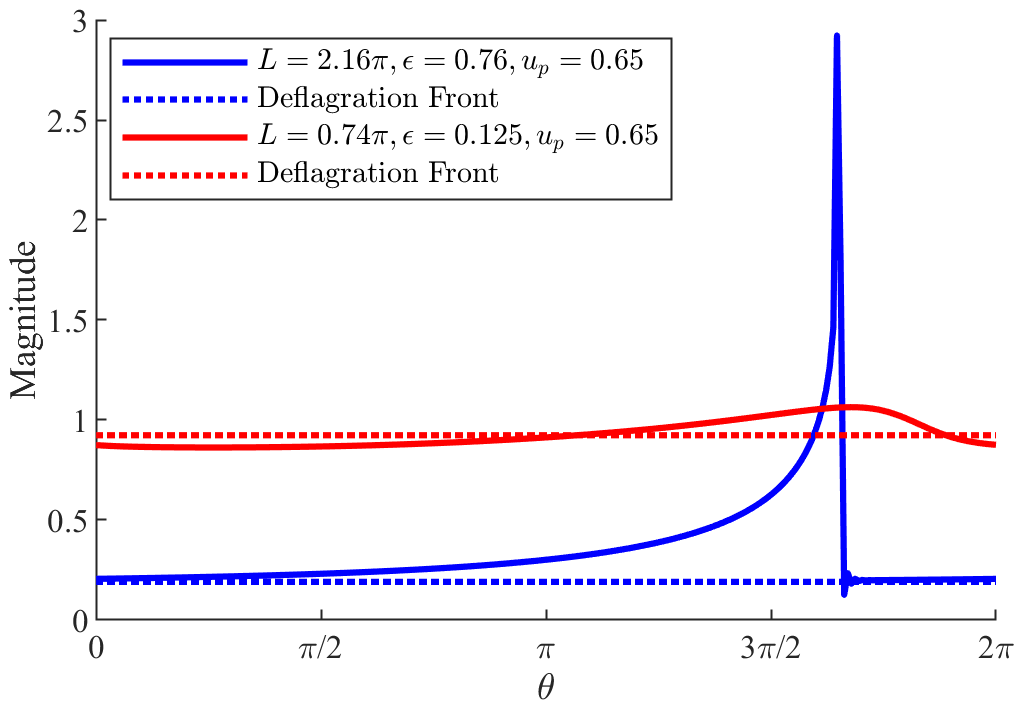} 
           
	    \end{overpic}  
	    \caption{Two traveling wave profiles corresponding to the diamond markers in Fig. \ref{fig:performance}. In the limit of large dissipation, the detonation wave beginning and end states are exactly equal to the magnitude of the deflagration branch. Implied is a loss of mode-locking properties, since the pulse is now isolated from its environment. The displayed weak wave has a negligible difference in  $\overline{u^2}$ compared to that of the associated deflagration state.}
		\label{fig:maxWave}
\end{figure}

To mimic such an area constriction or expansion in the RDE analog, the loss coefficient can be modified. In Fig. \ref{fig:performance}, displayed are several modeled systems with parameters as listed in Table \ref{tab:simulations}. Fixing the chemical potential of the modeled fluid ($q=1$) and the injection sensitivity threshold ($u_p = 0.65$), the domain length and loss coefficient $\epsilon$ are varied. In (a), these single-wave branches are displayed along side the deflagration branch for the system. As the loss coefficient is reduced (corresponding to a weak pressure gradient, perhaps because of a long flowpath or the addition of a geometric choke), the traveling wave branches merge with the deflagration branch: there is no discernible difference in $\overline{u^2}$. Physically, by increasing chamber pressure, the time scale associated with the kinetics becomes increasingly small, meaning deflagration is promoted in the entirety of the domain. The percent difference of $\overline{u^2}$ between the traveling wave and deflagration branches are displayed in (b). The domain lengths ($L=0.72\pi$ and $L=2.16\pi$) were empirically determined to be the shortest and longest (respectively) domain lengths to support a single traveling wave for the chosen model parameters. For the shortest round-trip distance (domain length of $L = 0.782\pi$), $\overline{u^2}$ is the highest, but there is a negligible difference between it and a planar deflagration. For the longest round-trip distance (domain length of $L = 2.16\pi$), $\overline{u^2}$ can attain the lowest value of the displayed systems, but the magnitude of  $\overline{u^2}$ is nearly 70\% higher than that of the planar deflagration.  A trade-off therefore exists: for propulsion systems, a general goal is to increase chamber pressure ($\overline{u^2}$ as presented here). However, this corresponds to \textit{decreasing} $\epsilon$, which in turn limits the strength of the waves, or in the extreme limit, eliminates all traveling waves. To increase dissipation corresponds to \textit{decreasing} chamber pressure, as one might expect. To do so would create a decidedly worse-performing engine than a deflagration-based system. But curiously, the chamber pressure for the rotating detonation mode can nearly double that of the trivial deflagration mode - a significant improvement can be had for these specific conditions.

The curves associated with traveling waves in Fig. \ref{fig:performance} terminate for increasing $\epsilon$. At the terminus marks the condition where dissipation now is the dominant physical process in the chamber - beyond this point, the wave is no longer in communication with its tail and it loses its mode-locking properties. This is shown in Fig. \ref{fig:maxWave}. The traveling wave profile for $L=2.16\pi$ and $\epsilon = 0.76$ (corresponding to the blue diamond marker in Fig. \ref{fig:performance}) is plotted along with the associated planar deflagration front. The time required to dissipate the wave tail dissipates to the state of the deflagration branch is identical to the transit time of the detonation wave. To contrast the properties of this wave, a weak wave (corresponding to the red diamond marker in Fig. \ref{fig:performance}) is additionally plotted. The weaker wave travels at 50\% of the Chapman-Jouguet speed of the system while the stronger wave travels at 78\% of this benchmark speed. Although the base-to-peak amplitudes of the stronger wave is approximately 14 times greater than that of the weaker wave, their wave speeds are of the same order.

\section{Conclusion} \label{sec:conclusion}
Rotating detonation waves exhibit a remarkable set of properties, including mode-locking and modulation. In examining the kinematic traces of the detonation waves, they are observed to undergo nonlinear interactions characteristic of solitons, including phase shifting and the exchanging of amplitude. The canonical solitonic structure of nonlinear waves is held together by the balance of nonlinearity and dispersion, subject to either Hamiltonian dynamics (the Kortewag-de Vries equation \citep{Gardner1971}, for example) or local gain-loss dynamics (passively mode-locked lasers \citep{Grelu2012}, for example). Unlike solitons of either Hamiltonian or canonical driven-dissipative systems, rotating detonation waves are held together through \textit{global gain dynamics}, where multi-scale physics associated with the unit processes of injection and mixing, combustion, exhaust, and wave propagation are all coupled and nonlinearly interact.

The rotating detonation analog system has been shown to adequately model these unit processes and their associated time scales. The model is successful in reproducing the solitonic structures and behaviors seen in experiments. Furthermore, we conclude that the behavior stemming from these multi-scale physics is \textit{unique} and \textit{fundamental} to the rotating detonation engine. With this study, we evaluate (linear) stability of modeled rotating detonation waves for the first time. We find that the steadily propagating pulse train undergoes a Hopf bifurcation to time-periodic modulation - this bifurcation to modulation we term the \textit{fundamental instability} of rotating detonation waves, as this is the bifurcation from which transient phenomena originate. 

Lastly, we acknowledge the engineering tasks at hand for maturing the rotating detonation engine technology. Most notably is the matching of time scales for the physical processes controllable by engineers: injection, mixing, and exhaust processes.

\begin{acknowledgments}
The authors acknowledge sponsorship under the U.S. Air Force Office of Scientific Research (AFOSR) Grant No. FA9550-18-1-9-007 and Office of Naval Research Funding Document No. N0001417MP00398. JNK acknowledge support from the Air Force Office of Scientific Research (AFOSR grant FA9550-17-1-0329).
\end{acknowledgments}

~\\
~\\
\appendix

\section{Continuation Formulation and Hopf Constraints} \label{app:hopf}
The continuation software \textit{pde2path} uses the finite element method to analyze partial differential equations of the form \citep{pde2path}
\begin{equation}
\frac{\partial u}{\partial t} = -G\left(u,\mathbf{\mu}\right)
\end{equation}
\begin{equation}
G\left(u,\mathbf{\mu}\right) = -\nabla\cdot\left(c\otimes\nabla u \right) + au -b\otimes\nabla u - f
\end{equation}
where $u$ is a function of space and time, $\mu$ is a vector of parameters, and the variables $a$,$b$,$c$ correspond to the linear, advection, and diffusion tensors, and $f$ is the nonlinearity. In the finite element formulation on a discretized domain, this reads as:

\begin{equation}
M\dot{u} = -G\left(u,\mathbf{\mu}\right)
\end{equation}
\begin{equation}
G\left(u,\mathbf{\mu}\right) = Ku - Mf\left(u,\mathbf{\mu}\right)
\end{equation}
where $M$ is the mass matrix, $K$ is the stiffness matrix (diffusion term),  and $Mf$ is the nonlinearity. The model exists on a periodic domain, meaning that a continuous symmetry exists that must be eliminated before attempting continuation. This is achieved by augmenting the system with an additional bifurcation parameter, $v$, corresponding to velocity, such that this \textit{imposed} speed exactly offsets the motion of the waves \citep{Uecker2019}. 

Therefore, the model system \ref{eq:cU}-\ref{eq:cL} is recast as:

\begin{multline} \label{eq:invariant}
G = \begin{pmatrix} \nu_1 K -vK_x& 0 \\ 0 & \nu_2 K-vK_x \end{pmatrix} \begin{pmatrix} u \\ \lambda \end{pmatrix} - \begin{pmatrix} -\frac{1}{2} K_x u^2 \\ 0 \end{pmatrix} \\ -\begin{pmatrix} M & 0 \\ 0 & M \end{pmatrix} \begin{pmatrix} kq(1-\lambda)\exp{\left(\frac{u - u_c}{\alpha}\right)} - \epsilon u^2 \\ k(1-\lambda)\exp{\left(\frac{u - u_c}{\alpha}\right)} - \frac{su_p\lambda}{1 + \exp{\left(r\left(u - u_p\right)\right)}} \end{pmatrix}
\end{multline}

The continuous symmetry associated with the periodic boundaries has been removed by the addition of this phase constraint. We use wave profiles from the direct numerical simulations from Fig. \ref{fig:dnsBif} to initialize the traveling wave branches for continuation. 

To compute time-periodic orbits along branches emanating from a detected Hopf bifurcation, an additional constraint must be imposed to fix the translational invariance in time \citep{Uecker2019}. Here, the imposed velocity of Eq. \ref{eq:invariant} is redefined to be the average velocity over the period of oscillation of the Hopf orbit. By defining a reference profile (the steady traveling wave profile, for example), the profile at each time slice can be compared to this reference, producing a deviation of traveling wave speed. Over one period of modulation, the average of these deviations from the reference profile are forced to be zero. Thus, time-periodicity is enforced and the translational invariance (in time) is fixed with respect to a reference profile. 

\bibliographystyle{aipauth4-1}
\bibliography{solitons}

\end{document}